\documentclass{emulateapj}

\usepackage{psfig}
\usepackage{url}
\usepackage{amsmath}
\usepackage{rotating}

\newcommand{\hstlong}{\textit{Hubble Space Telescope}}

\newcommand{\Chandra}{{\em Chandra}}
\newcommand{\chandra}{\textit{Chandra}}
\newcommand{\chandralong}{\textit{Chandra X-ray Observatory}}
\newcommand{\xmmlong}{\textit{XMM-Newton}}
\newcommand{\xmm}{\textit{XMM}}

\newcommand{\nh}{\mbox{$N_H$}}

\newcommand{\kteff}{\mbox{$kT_{\rm eff}$}}
\newcommand{\rinfty}{\mbox{$R_{\infty}$}}
\newcommand{\rns}{\mbox{$R_{\rm NS}$}}
\newcommand{\mns}{\mbox{$M_{\rm NS}$}}

\newcommand{\nhtt}{\mbox{$N_{H,22}$}}
\newcommand{\chisq}{\mbox{$\chi^2$}}
\newcommand{\chisqnu}{\mbox{$\chi^2_\nu$}}
\newcommand{\deltachisq}{\mbox{$\Delta\chi^2$}}
\newcommand{\Chisq}[3]{$\chi^2_\nu$/dof (prob.) = {#1}/{#2} (#3)}
\newcommand{\Msun}{\mbox{$M_\odot$}}

\newcommand{\Lx}{\mbox{$L_{\rm X}$}}

\newcommand{\mr}{\mbox{\mns--\rns}}
\newcommand{\MofR}{\mbox{\mns(\rns)}}

\newcommand{\xray}{\mbox{X-ray}}

\newcommand{\simlt}{\mathrel{\hbox{\rlap{\hbox{\lower4pt\hbox{$\sim$}}}\hbox{$<$}}}}
\newcommand{\simgt}{\mathrel{\hbox{\rlap{\hbox{\lower4pt\hbox{$\sim$}}}\hbox{$>$}}}}
\newcommand{\approxgt}{\mbox{$\,^{>}\hspace{-0.24cm}_{\sim}\,$}}
\newcommand{\approxlt}{\mbox{$\,^{<}\hspace{-0.24cm}_{\sim}\,$}}
\newcommand{\ee}[1]{\mbox{$10^{#1}$}}
\newcommand{\tee}[1]{\mbox{$\times 10^{#1}$}}
\newcommand{\ud}[2]{\mbox{$^{+ #1}_{- #2}$}}
\newcommand{\ppm}{\mbox{$\pm$}}

\newcommand{\unit}[1]{\mbox{$\rm\,#1$}}

\def\arcdeg{\hbox{$^\circ$}}
\def\arcmin{\hbox{$^\prime$}}
\def\arcsec{\hbox{$^{\prime\prime}$}}
\def\sec{\mbox{$\,{\rm sec}$}}
\newcommand{\msun}{\mbox{$\,M_\odot$}}

\newcommand{\km}{\hbox{$\,{\rm km}$}}

\newcommand{\MeV}{\mbox{$\,{\rm MeV}$}}
\newcommand{\keV}{\mbox{$\,{\rm keV}$}}

\newcommand{\ksec}{\mbox{$\,{\rm ks}$}}

\newcommand{\kpc}{\mbox{$\,{\rm kpc}$}}
\newcommand{\pc}{\mbox{$\,{\rm pc}$}}
\newcommand{\persec}{\mbox{$\,{\rm s^{-1}}$}}

\newcommand{\percmsq}{\mbox{$\,{\rm cm^{-2}}$}}
\newcommand{\percmcube}{\mbox{$\,{\rm cm^{-3}}$}}
\newcommand{\peryear}{\mbox{$\,{\rm yr^{-1}}$}}

\newcommand{\cgslum}{\mbox{$\,{\rm erg\,\persec}$}}
\newcommand{\cgsdensity}{\mbox{$\,{\rm g\percmcube}$}}

\newcommand{\PLunit}{\mbox{$\,{\rm keV^{-1}\persec\percmsq}$}}

\def\OmCen{\mbox{$\omega$\,Cen}}

\begin{document}

\title{Measurement of the Radius of Neutron Stars with High S/N
  Quiescent Low-mass \xray\ Binaries in Globular Clusters}

\author{Sebastien Guillot \footnote{Vanier Canada Graduate Scholar}} \affil{Department of
  Physics, McGill University,\\ 3600 rue University, Montreal, QC,
  Canada, H2X-3R4} \email{guillots@physics.mcgill.ca}

\author{Mathieu Servillat} \affil{Laboratoire AIM (CEA/DSM/IRFU/SAp,
  CNRS, Universit{\'e} Paris Diderot),\\ CEA Saclay, Bat. 709, 91191
  Gif-sur-Yvette, France} \affil{Harvard-Smithsonian Center for
  Astrophysics, 60 Garden Street, Cambridge, MA 02138, USA}

\author{Natalie A. Webb} \affil{Universit{\'e} de Toulouse; UPS-OMP; IRAP;
  Toulouse, France} \affil{CNRS; IRAP; 9 Av. colonel Roche, BP 44346,
  31028 Toulouse cedex 4, France}

\author{Robert E. Rutledge} \affil{Department of
  Physics, McGill University,\\ 3600 rue University, Montreal, QC,
  Canada, H2X-3R4} \email{rutledge@physics.mcgill.ca} 

\slugcomment{Accepted to ApJ}
\shorttitle{The Radius of Neutron Stars}

\begin{abstract}
This paper presents the measurement of the neutron star (NS) radius
using the thermal spectra from quiescent low-mass \xray\ binaries
(qLMXBs) inside globular clusters (GCs).  Recent observations of NSs
have presented evidence that cold ultra dense matter -- present in the
core of NSs -- is best described by ``normal matter'' equations of
state (EoSs).  Such EoSs predict that the radii of NSs, \rns, are
quasi-constant (within measurement errors, of $\sim10\%$) for
astrophysically relevant masses ($\mns>0.5\msun$).  The present work
adopts this theoretical prediction as an assumption, and uses it to
constrain a single \rns\ value from five qLMXB targets with available
high signal-to-noise X-ray spectroscopic data.  Employing a
Markov-Chain Monte-Carlo approach, we produce the marginalized
posterior distribution for \rns, constrained to be the same value for
all five NSs in the sample.  An effort was made to include all
quantifiable sources of uncertainty into the uncertainty of the quoted
radius measurement.  These include the uncertainties in the distances
to the GCs, the uncertainties due to the Galactic absorption in the
direction of the GCs, and the possibility of a hard power-law spectral
component for count excesses at high photon energy, which are observed
in some qLMXBs in the Galactic plane.  Using conservative assumptions,
we found that the radius, common to the five qLMXBs and constant for a
wide range of masses, lies in the low range of possible NS radii,
$\rns=9.1\ud{1.3}{1.5}\km$ (90\%-confidence).  Such a value is
consistent with low-\rns\ equations of state.  We compare this result
with previous radius measurements of NSs from various analyses of
different types of systems.  In addition, we compare the spectral
analyses of individual qLMXBs to previous works.

\keywords{stars: neutron --- X-rays: binaries --- globular clusters:
individual (\OmCen, M13, M28, NGC~6397, NGC~6304)}
\end{abstract}

\maketitle




\section{Introduction}
The relation between pressure and energy density in matter at and
above the nuclear saturation density $\rho_{c}=2.8\tee{14}
\cgsdensity$ is largely unknown \citep{lattimer01, lattimer07}.  This
is mostly due to uncertainties of many-body interactions as well as
the unknown nature of strong interactions and symmetry energy.  Inside
neutron stars (NSs), the equation of state of dense matter
($P\left(\epsilon\right)$, written dEoS, hereafter) can be mapped into
a mass-radius relation \MofR\ by solving the Tolman-Oppenheimer-Volkoff
equation \citep{oppenheimer39,misner73}.  Historically, well before
any observational constraints could be placed on the dEoS, nuclear
theory attempted to determine the $P\left(\epsilon\right)$ relation
that would govern the behavior of cold ultra-dense matter.  Since the
cores of NSs are composed of such matter, its behavior is of
astrophysical interest; likewise, the behavior of NSs due to the
composition of its core is of nuclear physics interest.

Three main families of dEoSs have been discussed in the last 10--20
years.  The first one regroups ``normal'' dense matter EoSs.  At
densities at $\rho_{c}$, nuclei dissolve and merge, leaving
undifferentiated nuclear matter in $\beta$-equilibrium.  In this type
of matter, the pressure is neutron-dominated via the strong force,
with a small proton fraction.  In other words, NSs are pressure
supported against gravity by neutron degeneracy.  The ``normal'' dEoSs
are calculated with a relativistic treatment of nucleon-nucleon
interactions, leading to a relation between pressure and density, with
the pressure vanishing at zero densities \citep{lattimer01}.  For NSs,
such dEoSs correspond to \MofR\ lines composed of two parts.  One
corresponds to constant low \mns\ at large \rns\ values. Then, as the
density increases, the \MofR\ relation for ``normal'' dEoSs evolves to
quasi-constant\footnote{Here, and elsewhere, we use the term
  ``quasi-constant'' to mean constant within measurement precision,
  $\sim10\%$.  This should be differentiated from a value which is
  constant when measured with infinite precision, or a value which is
  constant according to theory.} \rns\ as \mns\ increases, up to a
maximum mass, above which the NS collapses to a black-hole.  Examples
of the proposed form of these dEoSs include AP3-AP4 \citep{akmal97},
ENG \citep{engvik96}, MPA1 \citep{muther87}, MS0 and MS2
\citep{muller96}, and LS \citep{lattimer91}.

A second family of dEoSs is characterized by matter in which a
significant amount of softening (i.e., less pressure) is included at
high densities, due usually to a phase transition at a critical
density which introduces an additional hadronic or pure-quark
component in what is referred to as the NS's ``inner core''.
Additional components, such as a population of hyperons at large
densities (GM3, \citealt{glendenning91}), or kaon condensates (GS1,
GS2, \citealt{glendenning99}), have been considered.  For that reason,
these dEoSs are referred to as ``hybrid'' dense matter.
Because of this phase-transition, the maximum \mns\ is rather low
($\mns<1.7\msun$).  This also results in \MofR\ curves with a smooth
decrease in \mns\ from the maximum to the minimum \mns, as
\rns\ increases.  Some of the ``hybrid'' dEoSs are MS1
\citep{muller96}, FSU \citep{shen10a,shen10b}, GM3
\citep{glendenning91}, GS1 \citep{glendenning99}, and PAL6
\citep{prakash88}.

The third family of dEoSs relies on the assumption that strange quarks
compose matter in its ground state.  One characteristic of such matter
is that the pressure vanishes at a non-zero density, compared to the
other types of matter described above -- they have solid surfaces.  In
\mr\ space, these quark star dEoSs follow lines of increasing
\mns\ with increasing radius, up to a maximum radius.  Above this
value, \rns\ starts decreasing as \mns\ increases until \mns\ reaches
its own maximum, where the object collapses to a black-hole.  The
maximum \rns\ varies between $\sim 9\km$ and $\sim 11\km$, depending
on the model parameters used, namely, the strange quark mass $m_{s}$
and the quantum chromodynamic coupling $\alpha_{c}$ \citep{prakash95}.
Note that ``hybrid'' and ``normal'' matter stars do not have this
constraint, and their radii can theoretically be as large as
$\sim100\km$, at masses $\mns<0.5\msun$ \citep{lattimer01}

Since matter at such densities cannot be produced in Earth
laboratories, constraints on the dEoS theoretical models can only be
placed by the study of NSs, the only objects in the Universe
containing matter at such densities.  The measurements of \mns\ and
\rns\ have the potential to provide great insight to the theory of
cold ultra dense matter.  Various methods exist to measure \mns\ and
\rns\ \citep[e.g.,][for a general review]{lattimer07}.  These include
the study of quasi-periodic oscillations in active X-ray binaries
\citep{miller98,mendez07}, Keplerian parameters in NS binaries
\citep[][for \mns\ measurements]{nice04,demorest10}, thermonuclear
X-ray bursts \citep[][for \mr\ measurements]{ozel06,suleimanov11a},
pulse-timing analysis of millisecond pulsars
\citep{bogdanov08,bogdanov12}, and the thermal spectra of quiescent
low-mass X-ray binaries (qLMXBs), which is the method of this
investigation. Each of these different methods have their own unique
systematic uncertainties, and it is therefore of value to pursue each,
to permit intercomparison of their conclusions.

By itself, a \mns\ measurement can only place new constraints on the
dEoS when the measured value is above that of all previous
\mns\ measurements.  In \mr\ space, each dEoS predicts a maximum \mns,
above which the NS collapses to a black hole.  In particular, hybrid
dEoSs are characterized by a relatively low maximum
\mns\ ($\mns<1.8\msun$, \citealt{lattimer01}), while normal matter
dEoSs produce maximum \mns\ of up to 2.5\msun\ \citep{lattimer01}.
The maximum \mns\ for strange quark matter (SQM) dEoSs is typically in
the vicinity of 2\msun\ \citep{lattimer01}.  The maximum
\mns\ property of EoSs can be used to excludes dEoSs.  Historically,
\mns\ measurements were in the 1.3--1.5\msun\ range.  While the first
precise \mns\ measurements confirmed theoretical predictions about NSs
\citep[e.g.,][]{taylor89}, subsequent measurements at and below
previous values did not place any new constraints on the dEoS.
Recently, the mass of the radio pulsar PSR~1614$-$2230 was precisely
measured with a value $\mns=1.97\pm0.04\msun$ \citep{demorest10}.  The
implications of this measurement for nuclear physics have been
discussed with some depth \cite{lattimer11}.  Such a high
\mns\ excludes previously published hybrid models of dEoSs (using
specific values of assumed parameters from within their allowed
regions), although it does not rule out any specific form of exotica.
SQM dEoSs also seem to be disfavored, since their predicted maximum
\mns\ approaches the 2\msun\ limit for only some of the models within
the parameter spaces permitted by nuclear physics constraints.
Nonetheless, fine tuning of models may allow these disfavored dEoSs to
be marginally consistent with the \mns\ measured in PSR~1614$-$2230
(for example, \citealt{bednarek11,weissenborn12}, for hybrid models,
and \citealt{lai11}, for SQM models).

Overall, this high-\mns\ measurement seems to favor ``normal matter''
hadronic dEoSs. This would mean that the radius of astrophysical NSs
should be observed to be within a narrow ($\approxlt10\%$) range of
values for $\mns>0.5\msun$, since ``normal matter'' dEoSs follow lines
of quasi-constant radius in \mr-space at such masses
\citep{lattimer01}.  It is important to notice that the spread in
\rns\ increases for stiff EoSs, especially close to the maximum
\mns\ of the compact object (e.g., up to a 2-km difference in
\rns\ for the EoS PAL1, \citealt{prakash88}).

The empirical dEoS obtained from \mr\ confidence regions from type-I
\xray\ bursts and from the thermal spectra of qLMXBs combined also
favors this conclusion \citep{steiner10,steiner12}.  Using a Bayesian
approach, the most probable dEoS was calculated, resulting in a dEoS
approaching the behavior of theoretical hadronic dEoSs, with predicted
radii in the range $\rns\sim 10-13\km$.  Such radii suggest that soft
hadronic dEoSs are describing the dense matter inside NSs.  However,
different analyses of other NSs found radii consistent with stiff
dEoSs.  These include the qLMXB X7 in 47Tuc \citep{heinke06a}, or the
type I \xray\ burster 4U~1724-307 \citep{suleimanov11b}.  Nonetheless,
these results are not inconsistent with the observation that the
\rns\ is almost constant for a large range of \mns, since they are
consistent with stiff ``normal matter'' dEoSs, such as MS0/2
\citep{muller96}

Given the evidence supporting the ``normal matter'' hadronic dEoSs, it
therefore becomes a natural assumption -- to be tested against data --
that observed NSs have radii which occupy only a small range of
\rns\ values ($\approxlt 10\%$).  Using the thermal spectra of five
qLMXBs, fitted with a H-atmosphere model, a single \rns\ value is
assumed and measured, as well as its uncertainty. Furthermore, under
this assumption, the best-fit \mns\ and surface effective temperature
\kteff\ for these qLMXBs and their uncertainties are extracted.  The
various sources of uncertainty involved in this spectral analysis are
addressed, including, the distances to the qLMXBs, the amount of
galactic absorption in their direction, and the possibility of an
excess of high-energy photons as observed for other qLMXBs (and
modeled with a power-law component, PL hereafter).  The goal is to
place the best possible constraints on \rns\, accounting for all know
uncertainties, and eliminating all unquantifiable systematic
uncertainties.

\begin{deluxetable*}{lrrrrrrr}
  \tablecaption{\label{tab:Obs} \xray\ Exposures of the Targeted Clusters.}
  \tablewidth{0pt}
  \tabletypesize{\scriptsize}    
  \tablecolumns{8}
  \tablehead{
    \colhead{Target} & \colhead{Obs. ID} & \colhead{Starting} & \colhead{Usable time } &
    \colhead{S/N} & \colhead{Telescope} & \colhead{Filter} & \colhead{Refs.} \\
    \colhead{} &  \colhead{} & \colhead{Time (TT)} & \colhead{(ksec)} &
    \colhead{} & \colhead{and detector} & \colhead{or Mode}  & \colhead{}  }
  \startdata
  M28 & 2683 & 2002 July 04 18:02:19 &  14.0 & 23.85 & \chandra\ ACIS-S3 (BI) &  VFAINT & 2\\
  M28 & 2684 & 2002 Aug. 04 23:46:25 &  13.9 & 23.54 & \chandra\ ACIS-S3 (BI) &  VFAINT & 2\\
  M28 & 2685 & 2002 Sep. 09 16:55:03 &  14.3 & 23.90 & \chandra\ ACIS-S3 (BI) &  VFAINT & 2\\
  M28 & 9132 & 2008 Aug. 07 20:45:43 & 144.4 & 78.75 & \chandra\ ACIS-S3 (BI) &  VFAINT & 1, 3 \\
  M28 & 9133 & 2008 Aug. 10 23:50:24 &  55.2 & 48.46 & \chandra\ ACIS-S3 (BI) &  VFAINT & 1, 3 \\
  \hline
  NGC~6397 &  79  & 2000 July 31 15:31:33 &  48.34 & 25.03 & \chandra\ ACIS-I3 (FI) &  FAINT & 4, 5\\
  NGC~6397 & 2668 & 2002  May 13 19:17:40 &  28.10 & 25.47 & \chandra\ ACIS-S3 (BI) &  FAINT & 5\\
  NGC~6397 & 2669 & 2002  May 15 18:53:27 &  26.66 & 24.97 & \chandra\ ACIS-S3 (BI) &  FAINT & 5\\
  NGC~6397 & 7460 & 2007 July 16 06:21:36 & 149.61 & 52.31 & \chandra\ ACIS-S3 (BI) & VFAINT & 5\\
  NGC~6397 & 7461 & 2007 June 22 21:44:15 &  87.87 & 41.40 & \chandra\ ACIS-S3 (BI) & VFAINT & 5\\
  \hline
  M13 & 0085280301 & 2002 Jan. 28 01:52:41 & 18.8 & 14.25 & \xmm\ pn, MOS1, MOS2 & Medium & 6,7\\
  M13 & 0085280801 & 2002 Jan. 30 02:21:33 & 17.2 & 12.10 & \xmm\ pn, MOS1, MOS2 & Medium & 6,7\\
  M13 & 5436       & 2006 Mar. 11 06:19:34 & 27.1 & 16.07 & \chandra\ ACIS-S3 (BI) &  FAINT & 1,8 \\
  M13 & 7290       & 2006 Mar. 09 23:01:13 & 28.2 & 16.01 & \chandra\ ACIS-S3 (BI) &  FAINT & 1,8 \\
 \hline
  \OmCen & 653        & 2000 Jan. 24 02:13:28 & 25.3 & 13.33 & \chandra\ ACIS-I3 (FI) &  VFAINT & 9\\
  \OmCen & 1519       & 2000 Jan. 25 04:32:36 & 44.1 & 16.45 & \chandra\ ACIS-I3 (FI) &  VFAINT & 9\\
  \OmCen & 0112220101 & 2001 Aug. 12 23:34:44 & 33.9 & 24.35 & \xmm\ pn, MOS1, MOS2 & Medium & 7,10\\
  \hline
  NGC~6304 & 11074    & 2010 July 31 15:31:33 & 98.7 & 27.94 & \chandra\ ACIS-I3 (FI) & VFAINT & 1\\
  \enddata 

  \tablecomments{TT refers to Terrestrial Time.  FI and BI refers to
    the front-illuminated and back-illuminated ACIS chips. References:
    (1) This work; (2) \cite{becker03}; (3) \cite{servillat12}; (4)
    \cite{grindlay01b}; (5) \cite{guillot11a}; (6) \cite{gendre03b};
    (7) \cite{webb07}; (8) \cite{catuneanu13}, (9) \cite{rutledge02b};
    (10) \cite{gendre03a}; All observations have been re-processed and
    re-analyzed in this work.  The references provided here are given
    to indicate the previously published analyses of the data.}
\end{deluxetable*}

In this article, we provide the necessary theoretical background and
observational scenario to understand \mr\ measurements of NSs from
qLMXBs (\S~\ref{sec:qlmxb}).  The organization of the rest of the
paper is as follows: Section~\ref{sec:red} explains the analysis of
the \xray\ data. Section~\ref{sec:results} contains the results of the
spectral analysis. A discussion of the results is in
Section~\ref{sec:discussion} and a summary is provided in
Section~\ref{sec:conclusion}.

\section{Quiescent Low-Mass X-Ray Binaries}
\label{sec:qlmxb}

The low-luminosity of qLMXBs was initially observed following the
outbursts of the \xray\ transients Cen~X-4 and Aql~X-1
\citep{vanparadijs87}.  This faint emission
($\Lx\sim\ee{32}-\ee{33}\cgslum$, 4--5 orders of magnitude fainter
than during outburst) was originally interpreted as a thermal
blackbody emission.  Low-level mass accretion onto the compact object
was thought to explain the observed luminosity \citep{verbunt94}.

Later, an alternative to the low-level accretion hypothesis was
proposed.  This alternate theory, which became the dominant
explanation for the emission from qLMXBs, suggests that the observed
luminosity is provided, not by low-$\dot{M}$, but by the heat
deposited in the deep crust during outbursts \citep{brown98}.  In the
theory of deep crustal heating (DCH), the matter accreted during an
outburst releases $\sim 1.9\MeV$ of energy via pressure-sensitive
reactions: electron captures, neutron emissions or pycnonuclear
reactions \citep{sato79,haensel90,gupta07,haensel08}.  Therefore, the
time-averaged quiescent luminosity is proportional to the
time-averaged mass accretion rate:
\begin{equation}
  \langle L\rangle = 9\tee{32}\,\frac{\langle \dot{M}
    \rangle}{10^{-11}\unit{\msun\,
    \peryear}}\,\frac{Q}{1.5\unit{MeV/amu}}\cgslum
  \label{eq:dch}
\end{equation}
\noindent where $Q$ is the average heat deposited in the NS crust per
accreted nucleon \citep{brown98,brown00}.

Following this hypothesis about the energy source of the quiescent
luminosity, the theory of DCH also explains the observed spectra of
qLMXBs.  As a result of the energy deposited in the deep crust, the
core heats up during the outbursts.  The energy is then re-radiated on
core-cooling time scales away from the crust, and through the NS
atmosphere \citep{brown98}.  The NS atmosphere is assumed to be
composed of pure hydrogen.  Indeed, at the accretion rates expected
during quiescence, heavier elements settle on time scales of order
$\sim$\,seconds \citep{romani87, bildsten92}.  The possibility of
helium (He) or carbon atmospheres around NSs in LMXBs has also been
studied \citep{ho09,servillat12}.

Several models of H-atmosphere around NSs have been developed
\citep{rajagopal96, zavlin96, mcclintock04, heinke06a, haakonsen12}.
They are now routinely used to explain the emergent spectra of qLMXBs,
with emission area radii compatible with the entire surface area of
NSs, compared to derived emission area radii of $\approxlt 1\km$ in
the blackbody interpretation \citep{rutledge99}.  The DCH theory and
H-atmosphere models were first applied to explain the quiescent
spectra and measure the radius of historically transient LMXB (e.g.,
Cen~X-4, \citealt{campana00a, rutledge01b}, Aql~X-1,
\citealt{rutledge01a}).  However, the 10--50\% systematic uncertainty
on the distance to field LMXBs directly contributes to a 10--50\%
uncertainty on the radius measurements. Due to these large systematic
uncertainties, these objects provide limited use to place constraints
on the dEoS until more precise measurements of their distances can be
obtained.

Placing tight constraints on the dEoS requires $\sim5\%$ uncertainty
on the \rns\ measurements.  This constraint is approximately the
half-width of the range of radii in the \MofR\ relationships
corresponding to "normal matter" EoSs.  Globular clusters (GCs) have
properties which make them ideal targets for qLMXB observations:
relatively accurately measured distances; better characterized
Galactic absorption; over-abundances of LMXBs; and LMXBs with magnetic
field weak enough that the thermal spectrum is not affected
\citep{heinke06a}.  A handful of qLMXBs have been discovered in GCs so
far; only a few have \xray\ spectra with the high signal-to-noise
ratio (S/N) necessary to measure \rns\ with $\approxlt 10-15\%$
uncertainty, including the uncertainty to their distances.

\section{Data Reduction and Analysis}
\label{sec:red}

\begin{figure}[ht]
    \centerline{~\psfig{file=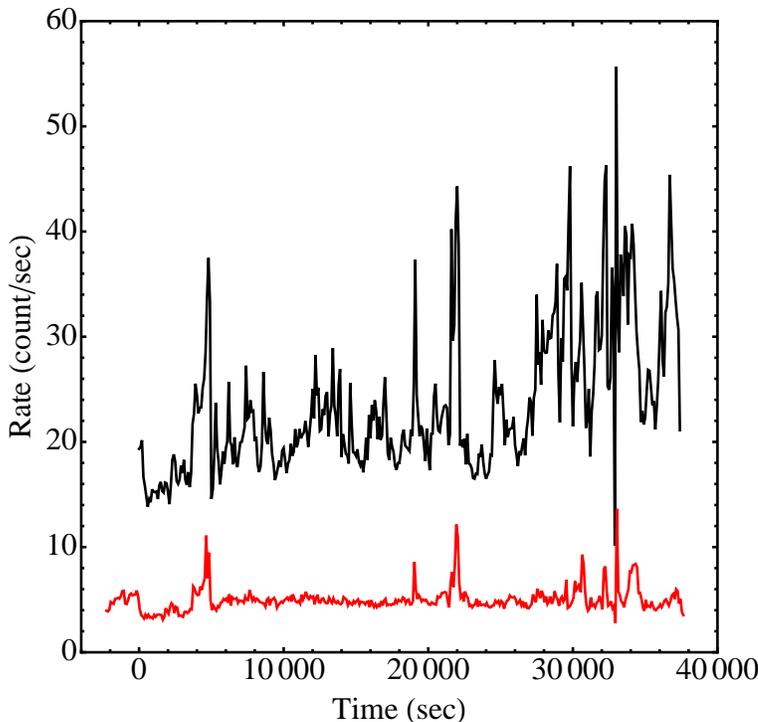,width=10cm,angle=0}~}
    \bigskip
    \caption[]{\label{fig:LC_OmCen} Figure showing the \xmm-pn
      full-detector light curves of \OmCen, ObsID 0112220101, with
      bins of 100\sec. The black (top) line corresponds to the pn
      camera light curve. The $t=0\sec$ time is the beginning of the
      pn exposure, on 2001 Aug. 12, 23:34:44.  The MOS1 light curve,
      in red (bottom), is shown for completeness and because periods
      of flaring are more readily visible.  The time intervals with
      large background flaring are excluded from the analyzed data
      set.}
\end{figure}

\subsection{Targets}
\label{sec:targets}

The targets used in this work are chosen among the qLMXBs located in
GCs that produced the best \rns\ measurements, i.e., with
\rinfty\ uncertainties of \approxlt 15\% in the previous works. 

The GCs \OmCen\ \citep{rutledge02b, gendre03a} and M13
\citep{gendre03b,catuneanu13} each have one qLMXB that was used in
previous work to place moderate constraints on the dEoS
\citep{webb07}.  The projected radius measurements \rinfty\ reported
in the original works are within 2--3\% uncertainty.  However, there
is evidence that these uncertainties are highly under-estimated
(\S~\ref{sec:results}).

The qLMXB in the core of NGC~6304, discovered recently with the
\xmmlong\ observatory (\xmm, hereafter) and confirmed with a short
\chandralong\ exposure \citep{guillot09a,guillot09b}, was then
observed for 100\ksec\ with ACIS-I onboard \chandra\ (Advanced
Charge-coupled-device Imaging Spectrometer).  In this work, only the
long \chandra\ exposure is used since, in the \xmm\ observation, the
core source is contaminated by nearby sources, mostly one spectrally
hard source \citep{guillot09b}.

The qLMXB in NGC~6397 (named U24 in the discovery observation,
\citealt{grindlay01b}) has a \rns\ value measured with $\sim 8\%$
uncertainty, obtained from a total of 350\ksec\ of \chandralong\ archived
observations \citep{guillot11a}.  The spectra for this target were
re-analyzed in this work, for a more uniform analysis.

Finally, the \rinfty\ measurement of the NS qLMXB in the core of M28
reported in the discovery observation does not place useful
constraints on the dEoS: $\rinfty=14.5\ud{6.9}{3.8}$ \citep{becker03}.
However, an additional 200\ksec\ of archived observations with
\chandra\ have been analyzed in a recent work, finding
$\rns=9\ppm3\km$ and $\mns=1.4\ud{0.4}{0.9}\msun$ for a H-atmosphere,
and $\rns=14\ud{3}{8}$ and $\mns=2.0\ud{0.5}{1.5}\msun$ for a pure
He-atmosphere \citep[][and their Figures 3 and 4, for the
  \mr\ confidence regions]{servillat12}.  The same data sets are used
in the present work.  This source is moderately piled-up ($\sim4\%$
pile-up fraction) and necessitates the inclusion of a pile-up model
component \citep[][see \S~\ref{sec:pileup} for details]{davis01}.  All
uncertainties for values obtained from a \xray\ spectral analysis with
\emph{XSPEC} are quoted at the 90\% confidence level, unless noted
otherwise.

The qLMXB X7 in 47~Tuc has also been observed with the high S/N that
could provide constraints on the dEoS.  However, it suffers from a
significant amount of pile-up (pile-up fraction $\sim 10-15\%$).
While the effects of pile-up can be estimated and corrected for by the
inclusion of a pile-up model \citep{heinke03a,heinke06a}, the
uncertainties involved with such a large amount of pile-up are not
quantified in this model\footnote{See "The Chandra ABC Guide to
  Pile-Up" available at
  \url{http://cxc.harvard.edu/ciao/download/doc/pileup\_abc.pdf}}.  It
was chosen not to include this target in the present analysis, in an
effort to limit the sources of uncertainties that are not quantified
(see \S~\ref{sec:pileup}).

The list of targets and their usable observations with \xmm\ and
\chandra\ is presented in Table~\ref{tab:Obs}, along with the usable
exposure time and other relevant parameters of the observations.

\subsection{Data Processing}
\label{sec:reduction}

The processing of raw data sets is performed according to the standard
reduction procedures, described briefly below.

\subsubsection{\chandralong\ Data Sets}
The reduction and analysis of \chandra\ data sets (ACIS-I or ACIS-S)
is done using {\tt CIAO~v4.4}.  The level-1 event files were first
reprocessed using the public script \emph{chandra\_repro} which
performs the steps recommended in the data preparation analysis
thread\footnote{\url{http://cxc.harvard.edu/ciao/threads/data.html}}
(charge transfer inefficiency corrections, destreaking, bad pixel
removal, etc, if needed) making use of the latest effective area maps,
quantum efficiency maps and gain maps of CALDB~v4.4.8
\citep{graessle07}.  The newly created level-2 event files are then
systematically checked for background flares.  Such flares were only
found in the middle and at the end of an observation of M28 (ObsID
2683), for a total of 3\ksec.  These two flares caused an increase by
a factor of 2.4, at most, of the background count level.  Given the
extraction radius chosen here (see below), this period of high
background contaminates the source region with $<1$ count.  Therefore,
the entire exposure of the ObsID is included in the present analysis.

To account for the uncertainties of the absolute flux calibration, we
add systematics to each spectral bins using the {\tt heasoft} tool {\tt
  grppha}.  In the 0.5--10\keV\ range, we add 3\% systematics
\citep[Table 2 in Chandra X-Ray Center Calibration Memo by
][]{edgar04}.  This document (Table 2) provides the uncertainties on
the ACIS detector quantum efficiency at various energies, which are
3\% at most.  In the 0.3--0.5\keV\ range, the uncertainty in the
calibration is affected by the molecular contamination affecting ACIS
observations.  The recent version of CALDB contains an improved model
for this contamination.  The RMS residuals are now limited to 10\% in
the 0.3--0.5\keV\ range (Figure 15 of document ``Update to ACIS
Contamination Model'', Jan 8, 2010, ACIS Calibration
Memo\footnote{available at
  \url{http://cxc.harvard.edu/cal/memos/contam\_memo.pdf}}).  To
account for the variations in the residuals of contamination model, we
add 10\% systematics to spectral bins below 0.5\keV.

\subsubsection{\xmmlong\ Data Sets}
The reduction of \xmm\ data sets is completed using the {\tt
  \xmm\ Science Analysis System v10.0.0} with standard procedures.
The command \emph{epchain} performs the preliminary data reduction and
creates the event files for the pn camera in the 0.4--10.0\keV\ energy
range, with $3\%$ systematic uncertainties included, accounting for
the uncertainties in the flux calibration of the pn camera on
\xmm\ \citep{guainizzi12}.  The data sets are checked for flares and
time intervals with large background flares are removed.  The total
usable time (after flare removal) for each observation is listed in
Table~\ref{tab:Obs}.  MOS1 and MOS2 data are not used in the present
work to minimize the effects of cross-calibration uncertainties
between detectors.

\begin{figure*}[ht]
  \centering
  \begin{tabular}{ccc}
    (a) & (b) & (c) \\
    \psfig{file=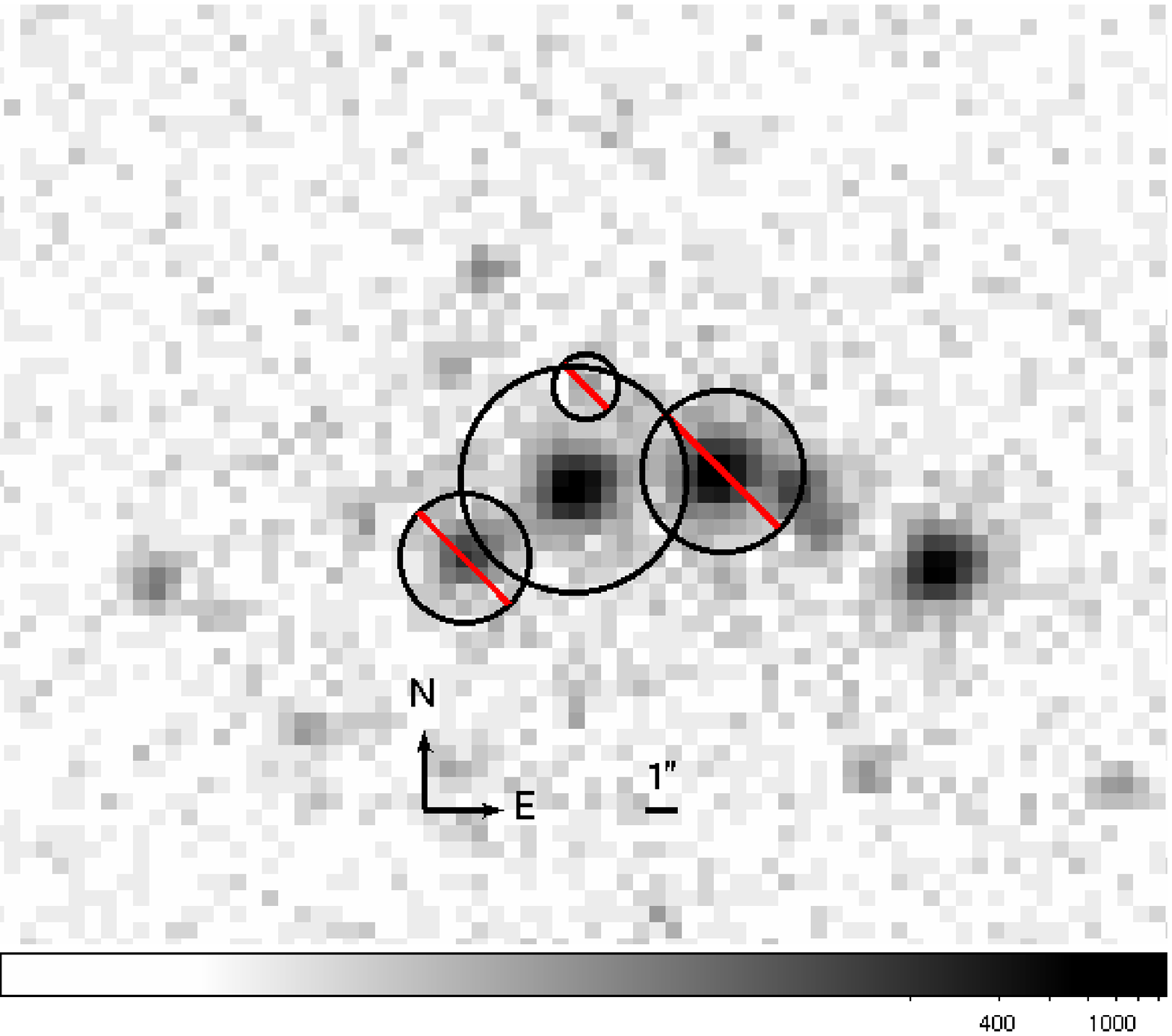,width=5.cm,angle=0} &
    \psfig{file=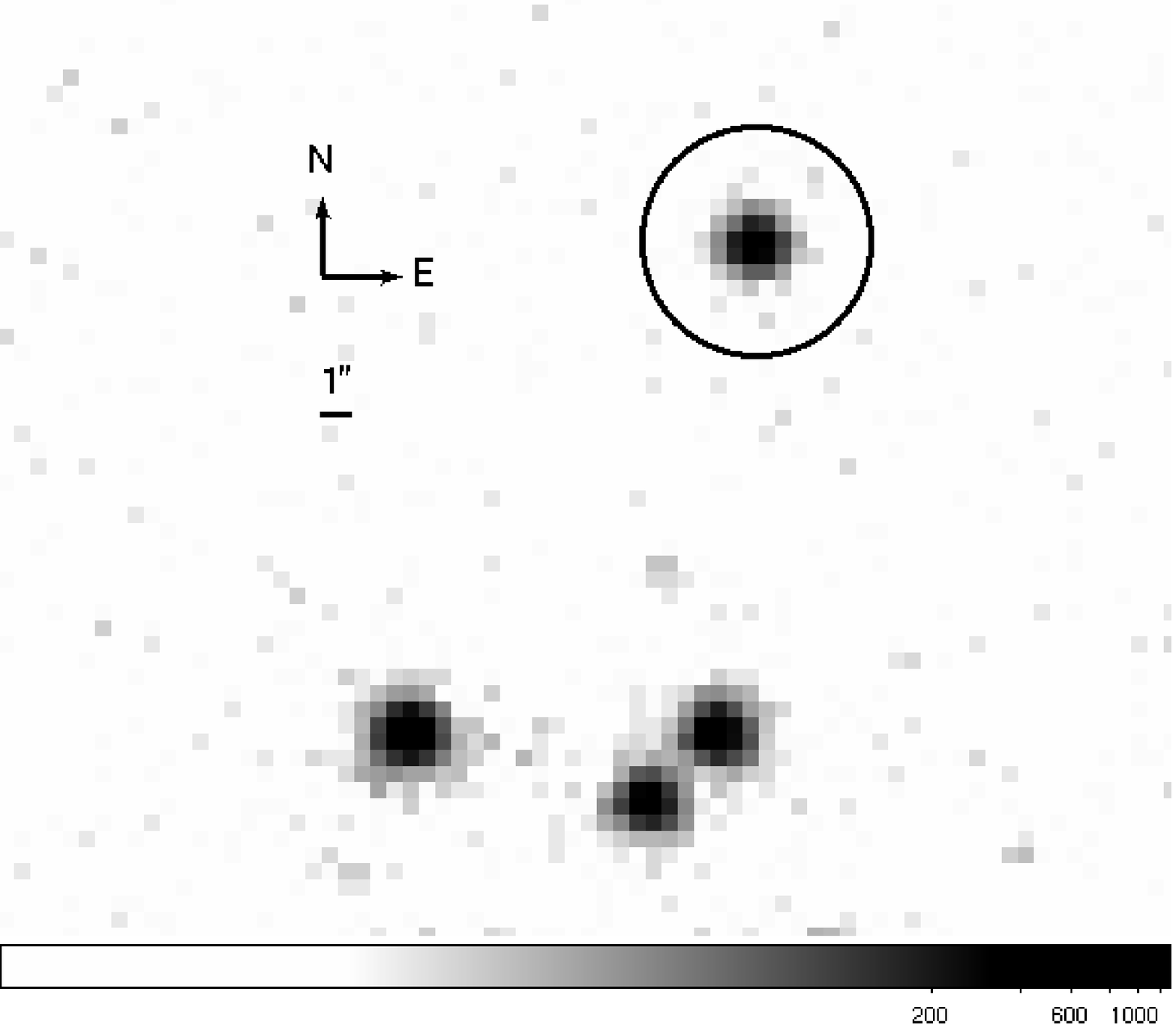,width=5.0cm,angle=0} &
    \psfig{file=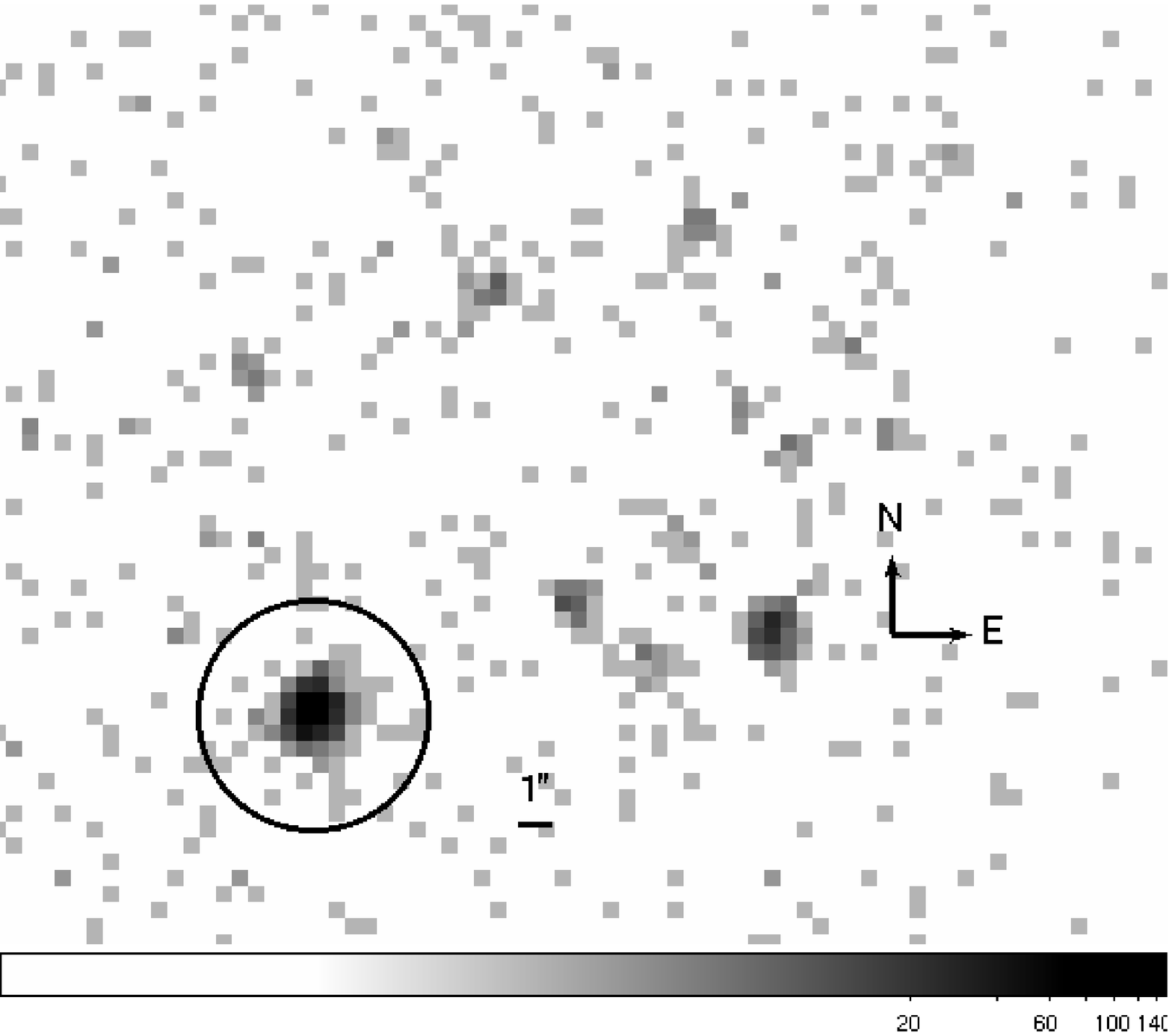,width=5.0cm,angle=0} \\
    \multicolumn{3}{c}{(d)} \\
    \multicolumn{3}{c}{\psfig{file=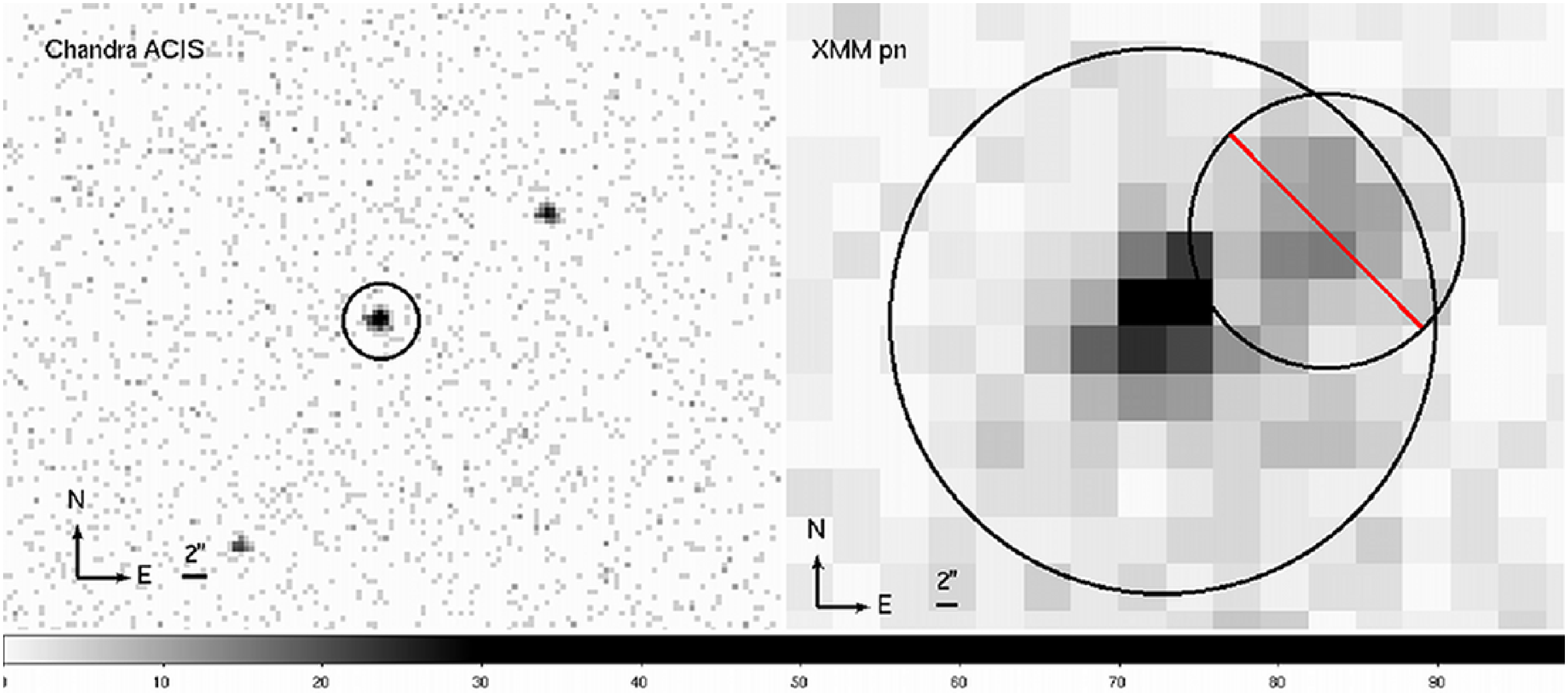,width=12cm,angle=0}} \\
    \multicolumn{3}{c}{(e)} \\
    \multicolumn{3}{c}{\psfig{file=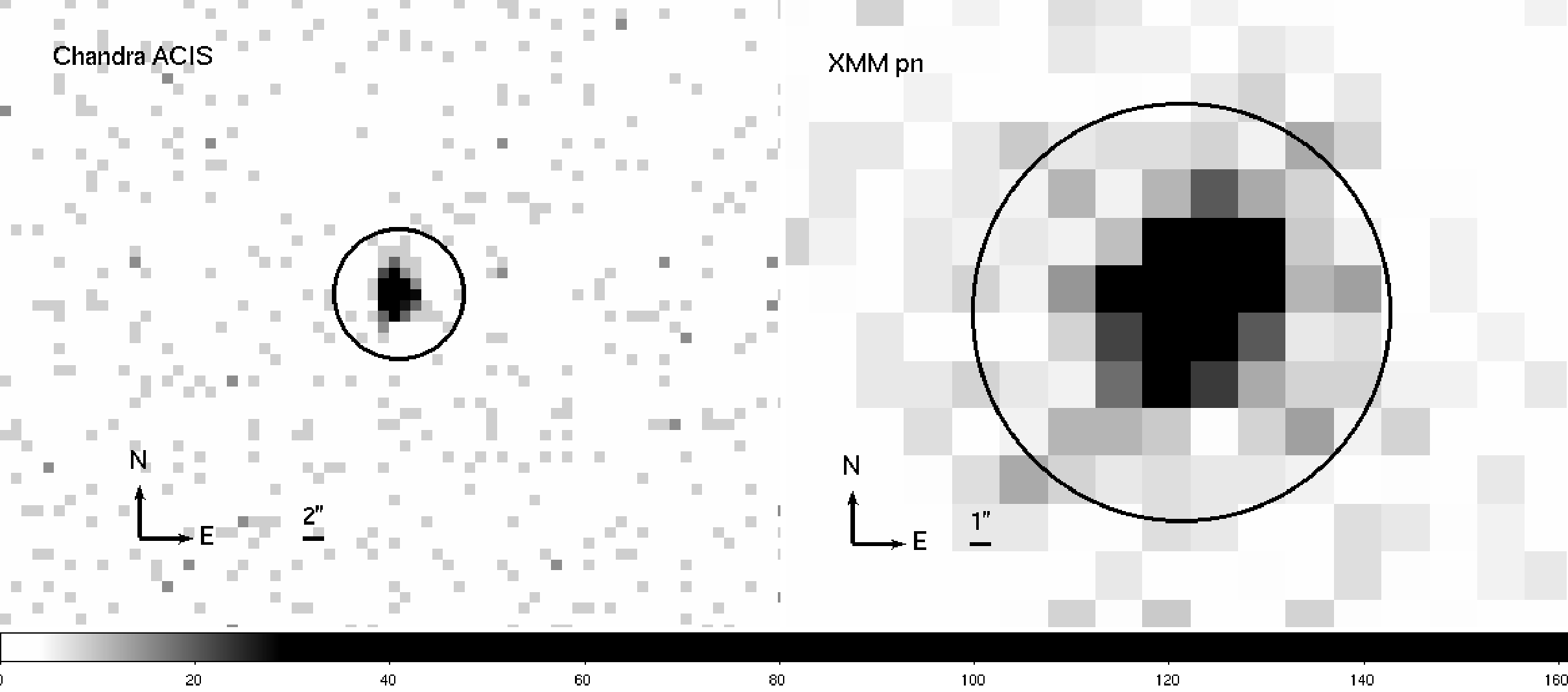,width=12cm,angle=0}} \\
  \end{tabular}
  \caption[]{\label{fig:regions} Figure showing the extraction
    regions for the five qLMXBs. {\it (a)} For M28, ObsID 9132, the
    three nearby sources are excluded, limiting contamination to
    $<1\%$ within the extraction region (see
    Table~\ref{tab:contamination}).  {\it (b)} For NGC~6397, ObsID
    7460, no counts from nearby sources fall within the extraction
    region. {\it (c)} For NGC~6304, ObsID 11074, the nearby sources
    are not contaminating the extraction region. {\it (d)} For M13,
    \chandra\ data are on the left, ObsID 7290, and \xmm\ data are on
    the right, ObsID 0085280301.  The nearby CV is excluded. {\it (e)}
    For \OmCen, \chandra\ data are on the left, ObsID 1519, and
    \xmm\ data are on the right, ObsID 0112220101.  There is no
    contamination from nearby sources.}
\end{figure*}

\subsection{Count Extraction}
\label{sec:extraction}

\subsubsection{On \chandralong\ Data Sets}
The count extraction of the source and background spectra is performed
with the task {\tt specextract}, as well as the calculation of the
response matrices and ancillary response files (RMFs and ARFs). The
centroid position is chosen using the reported source positions from
previous works.  The extraction radii are chosen to correspond to a
99\% EEF (Encircled Energy Fraction), and therefore depend on the
off-axis angle of the targets.  For on-axis qLMXBs (M13, M28, NGC~6397
and NGC~6304), counts within a 3.4\arcsec\ radius are extracted to
create the spectrum.  This ensures that 99\% of the enclosed energy
fraction at 1\keV\ is included \footnote{\Chandra\ Observatory Proposer
  Guide v15.0, fig.~6.7, December 2012}.  The qLMXB in \OmCen\ is at a
large off-axis angle ($\sim4.4\arcmin$) which requires an extraction
radius of 6\arcsec\ to contain 99\% of the EEF.  This is due to the
degradation of the PSF of the \chandra\ mirror with increasing
off-axis angle.  Background counts were taken from an annulus centered
around the qLMXB, with an inner radius of 5\arcsec\ (9\arcsec\ for
\OmCen) and an outer radius of 50\arcsec.  Regions surrounding other
point sources detected in the qLMXB extraction regions or in the
background regions are also excluded (radius of 5\arcsec\ or more).
For NGC~6304, the background region is off-centered with respect to
the source region, to ensure that the background lies on the same CCD
chip as the source.

Figures~\ref{fig:regions}a--e show the regions used to extract the
counts and create the spectra of each target. When several
observations are available for a target, the largest-S/N observation
was used to create the figure.

\subsubsection{On \xmmlong\ Data Sets}
For the three \xmm\ data sets, the extraction method was the same at
that described above. Only the source extraction radii were different
and determined using the XMM~SAS task {\tt eregionanalyse} which
provides the optimum extraction radius that maximizes the S/N given
the source position and the surrounding background.  The optimum
radius is 19\arcsec\ for ObsID 01122 of \OmCen.  The encircled energy
of the source is therefore 79\% at 1.5\keV.

For the \xmm\ observations of M13, the close proximity of a
cataclysmic variable (CV) complicates the task.  A
25\arcsec\ extraction radius and a 12.5\arcsec\ exclusion radius for
the nearby source are used to create the spectra.  It ensures that
84\% of the total energy from the qLMXB at 1.5\keV\ is
encircled\footnote{From \xmm\ Users Handbook, fig. 3.7, July
  2010}.

Similar to the \chandra\ data, the background is an annulus around the
source, restricted to remain on the same CCD chip as the source. {\tt
  rmfgen} and {\tt arfgen} are then used to generate the response
matrices files (RMF) and the ancillary response file (ARF) for each
observations.

\subsubsection{Contamination from Nearby Sources}
As mentioned above, some qLMXBs lie in close proximity of other
contaminating sources.  The most evident case is that of M13, observed
with \xmm\ (Figure~\ref{fig:regions}e), in which part of the counts
from a nearby CV still overlaps with the qLMXB extraction region even
after excluding 10\arcsec\ around the CV.  In M28, three sources in
the proximity of the qLMXB require parts of the extraction region to
be excluded (Figure~\ref{fig:regions}a), with a minor contamination
from the nearby sources.  For the observations of M28 and M13, the
fraction of contaminating counts present within the extraction region
of the qLMXBs is estimated using a Monte Carlo sampler which draws
counts from the radial distribution of encircled energy of ACIS or pn
at 1\keV\ (see footnotes in previous subsections).
Table~\ref{tab:contamination} lists the amount of contamination for
each observation of M28 and for the \xmm-pn observations of M13.  The
contamination over the M28 qLMXB region can be neglected, since, in
the worst case (ObsID 9132) only 47 counts out of 6250 are
contamination from nearby sources.  The CV close to the qLMXB in M13
causes a contamination of 6\%\ and 9\%\ of the counts in each of the
\xmm\ spectra, or 4\%\ of the total counts available for the qLMXB in
M13 (\xmm\ and \chandra\ spectra combined).  Overall, contamination
for nearby sources represents 0.5\%\ of the total count number, all
sources combined.  This contamination can be safely neglected, since
it will not significantly affect the radius measurement.

\subsection{Pile-Up}
\label{sec:pileup}
Observations of bright \xray\ sources may be subject to an
instrumental effect known as pile-up.  When two or more photons strike
a pixel on an \xray\ detector within a single time frame
(3.24\sec\ for \Chandra-ACIS observations and 73.4\unit{ms} for
\xmm-pn observations), the pile-up effect causes degradation of the
PSF and, more importantly for the analysis presented here, a
degradation of the spectral response.  Specifically, the recorded
energy of the event will be the sum of the two (or more) piled-up
photon energies.  In addition, grade migration (also called photon
pattern distortion for \xmm) also occurs.  Although a pile-up model
exists in \emph{XSPEC} to take into account these effects
\citep{davis01}, it is chosen here to restrict the analysis to mildly
piled-up observations.

None the \xmm\ observations of qLMXBs are piled-up, given the short
duration of a single time frame on \xmm-pn (73.4\unit{ms}).
Quantitatively, the count rates of the qLMXBs in \OmCen\ and M13
($2.6\tee{-2}$ and $2.7\tee{-2}$ counts per seconds, respectively)
correspond to $\sim\ee{-3}$ counts per frame.  At those rates, the
pile-up is negligible.  The frame time of \chandra-ACIS in full-frame
mode, however, is significantly longer than that of \xmm-pn
(compensated by the smaller effective area).  The
\chandra\ observations of the qLMXB in M28 are moderately piled-up
because of a count rate of $\sim0.043$ counts per seconds ($\sim0.14$
counts per frame) which corresponds to a pile-up fraction of
$\sim5\%$\footnote{\Chandra\ Observatory Proposer Guide v12.0,
  fig.~6.18, December 2009}.  Such amount of pile-up cannot be
neglected and is taken into account using the {\tt pileup} model in
\emph{XSPEC} \citep{davis01}.  Other \chandra\ observations of the
qLMXBs studied in this work have smaller count rates which do not
necessitate a pile-up correction.

As stated before, the correction of pile-up fractions $\sim 10\%$ and
above (such as that of the qLMXB in 47Tuc, observed in full-frame mode
on \chandra/ACIS) comes with unquantified uncertainties.  This is
tested by simulating piled-up spectra of 47Tuc (10--15\%\ pileup
fraction) and M28 (5\%\ pileup fraction) with their respective {\tt
  nsatmos} parameters, and then by fitting the spectra with the {\tt
  nsatmos} model without the pile-up component.  The best-fit radii of
each spectrum is affected by systematics: $\sim 10\%$ for M28 and
$\sim 50\%$ for 47Tuc.  The systematic error involved with the pile-up
of M28 is smaller than the measurement error of \rns\ in the present
analysis, while for 47Tuc, the systematic bias caused by pileup is
substantially larger than the \rns\ measurement uncertainty.
Therefore, the qLMXB in 47Tuc is not used in this analysis to avoid
introducing a systematic uncertainty which may be comparable in size
to our total statistical uncertainty \footnote{Note that we find in
  Section~\ref{sec:add47tuc} that inclusion of 47Tuc X7 in this
  analysis does not significantly affect our best-fit \rns\ value;
  nonetheless, we do not include this data, since we cannot estimate
  the effect of including it on our error region.}.

\subsection{Spectral Analysis}
\label{sec:analysis}

The spectral analysis is composed of two parts.  In the first one, the
five targets are analyzed individually and the results are compared to
previously published results.  The second part of the analysis in the
present work pertains to the simultaneous fitting of the targets, with
a \rns\ value common to all five qLMXBs.  Prior to the discussion of
these two parts, the analysis techniques common to the two analyses
are described.

\subsubsection{Counts Binning, Data Groups and Model Used}
\label{sec:model}

\begin{deluxetable}{crccccc}
  \tablecaption{\label{tab:contamination} Count Contamination from Nearby Sources}
  \tablewidth{0pt}
  \tabletypesize{\scriptsize}    
  \tablecolumns{7}
  \tablehead{
    \colhead{Target} & \colhead{ObsID} & \multicolumn{4}{c}{Number of
      contaminating counts} & \colhead{Total} \\ 
    \colhead{} & \colhead{(detector)} & \colhead{source 1} &
    \colhead{source 2} & \colhead{source 3} & \colhead{sum} &
    \colhead{(\%)}}
  \startdata
    M28      & 2693 (ACIS) & 0.16 & 0.11 & 0.16 & 0.43 & 0.08\% \\
             & 2684 (ACIS) & 0    & 0.15 & 0.32 & 0.47 & 0.08\% \\
             & 2685 (ACIS) & 0.05 & 0.19 & 0.17 & 0.37 & 0.06\% \\
             & 9132 (ACIS) & 1.7  & 38.9 & 6.3  & 46.9 & 0.75\% \\
             & 9133 (ACIS) & 1.0  & 11.7 & 2.7  & 15.4 & 0.65\% \\ 
     M13     & 0085280301 (pn) & 27 & -- & -- & 27 & 8.7\% \\
             & 0085280801 (pn) & 34 & -- & -- & 34 & 6.2\% \\
  \enddata
  \tablecomments{The columns ``source1'', ``source2'' and ``source3''
    indicate the absolute numbers of counts falling within the qLMXB
    extraction region, and ``sum'' is the simple sum of contaminating
    counts.  For M13, there is only one nearby source, not fully
    resolved with \xmm.  The last column provides the amount of
    contamination as a percent of the total number of counts within
    the qLMXB extraction region.}
\end{deluxetable}

Once the spectra and the respective response files of each observation
are extracted, the energy channels are grouped with a minimum of 20
counts per bin to ensure that the Gaussian approximation is valid in
each bin.  For observations with a large number of counts ($>$2000
counts, in ObsID 9132 and 9133 of M28 and ObsID 7460 of NGC~6397), the
binning is performed with a minimum of 40 counts per bin.  In all
cases, when the last bin (at high energy, up to 10\keV) contains less
than 20 counts, the events are merged into the previous bin.

The spectral fitting is performed using the ``data group'' feature of
\emph{XSPEC}.  The spectra of each target are grouped together, and
each group (corresponding to each qLMXB) is assigned the same set of
parameters.  The spectral model used is the {\tt nsatmos} model
\citep{heinke06a}, together with Galactic absorption taken into
account with the multiplicative model {\tt wabs}.  The amounts of
Galactic absorption, parameterized by \nh\ (\nhtt\ in units of
$\ee{22}\unit{atoms\percmsq}$, hereafter), are fitted during the
spectral analysis, and compared to those obtained from a HI Galactic
survey from NRAO data\footnote{obtained from the {\it HEASARC}
  \nh\ tools available at
  \url{http://heasarc.nasa.gov/cgi-bin/Tools/w3nh/w3nh.pl}}
\citep{dickey90}.  The \nh\ values used in the present analysis are
shown in Table~\ref{tab:GC}.  The results obtained with {\tt nsatmos}
are also compared with the best-fit results using the model {\tt
  nsagrav} \citep{zavlin96} for completeness.

As mentioned before, the {\tt pileup} model \citep{davis01} is
necessary for the spectral fitting of M28 spectra.  In \emph{XSPEC},
multiple groups cannot be fitted with different models, so a single
model is applied to all groups, namely, {\tt pileup*wabs*nsatmos}.
For the spectra of the qLMXB in M28, the $\alpha$ parameter of the
{\tt pileup} model, called ``good grade morphing parameter'' is left
free.  The frame time parameter is fixed at $3.10\sec$.  This value
corresponds to the {\tt TIMEDEL} parameter of the header
(3.14104\sec\ for the observations of M28) where the readout time
(41.04\unit{ms}) is subtracted. All the other parameters of the {\tt
  pileup} model are held fixed at their default values, as recommended
in the document "The Chandra ABC Guide to Pile-Up v.2.2"\footnote{from
  the Chandra \xray\ Science Center (June 2010), available at
  \url{http://cxc.harvard.edu/ciao/download/doc/pileup\_abc.pdf}}.
Since the targets in M13, \OmCen, NGC~6397 and NGC~6304 do not require
to account for pile-up, the time frame for these four groups is set to
a value small enough so that the {\tt pileup} model has essentially no
effect and the $\alpha$ parameters of the four non piled-up sources
are kept fixed at the default value, $\alpha=1$.  A quick test is
performed to demonstrate that the {\tt pileup} model with a small
frame time has no effect on the spectral fit using the non piled-up
spectra of the qLMXB in NGC~6304.  Specifically, the best-fit {\tt
  nsatmos} parameters and the \chisq-statistic do not change when the
{\tt pileup} model (with a frame time of 0.001\sec) is added, as
expected.

\subsubsection{Individual Targets}
Prior to the spectral analysis, it is important to verify that the
qLMXBs do not present signs of spectral variability.  This is done by
considering each target individually (without the other targets) and
by demonstrating that the {\tt nsatmos} spectral model fits
adequately, with the same parameters, all the observations of the
given target.  More precisely, for each target, all the parameters are
tied together and we verify that the fit is statistically acceptable.

In Section~\ref{sec:rinfty}, the results of the spectral fits of
individual targets are presented.  In order to provide the full
correlation matrix for all the parameters in the fit, a Markov-Chain
Monte-Carlo (MCMC) simulation is implemented (described in
\S~\ref{sec:mcmc}) and the resulting posterior distributions are
used as the best-fit confidence intervals, including the
\mr\ confidence region.  These spectral fitting simulations are
performed with the Galactic absorption parameters \nh\ left free.
This allows us to obtain best-fit \xray-measured values of the
absorption in the direction of each of the targeted GCs.  These
best-fit values are compared to HI-deduced values (from neutral
hydrogen surveys), and are also used for the remainder of the work,
when \nh\ is kept fixed.  Finally, the results of the individual
spectral fits are compared to previously published results.

\subsubsection{Simultaneous Spectral Analysis of the Five Targets}

The main goal of this paper is the simultaneous spectral fit of five
qLMXBs assuming an \rns\ common to all qLMXBs.  Therefore, \rns\ is a
free parameter constrained to be the same for all data sets, while
each NS targeted has its own free \mns\ and \kteff\ parameters.  In
addition, the spectra of M28 require an extra free parameter $\alpha$
for the modeling of pile-up.  This leads to a total of 12 free
parameters.

\begin{figure*}[ht]
  \centerline{~\psfig{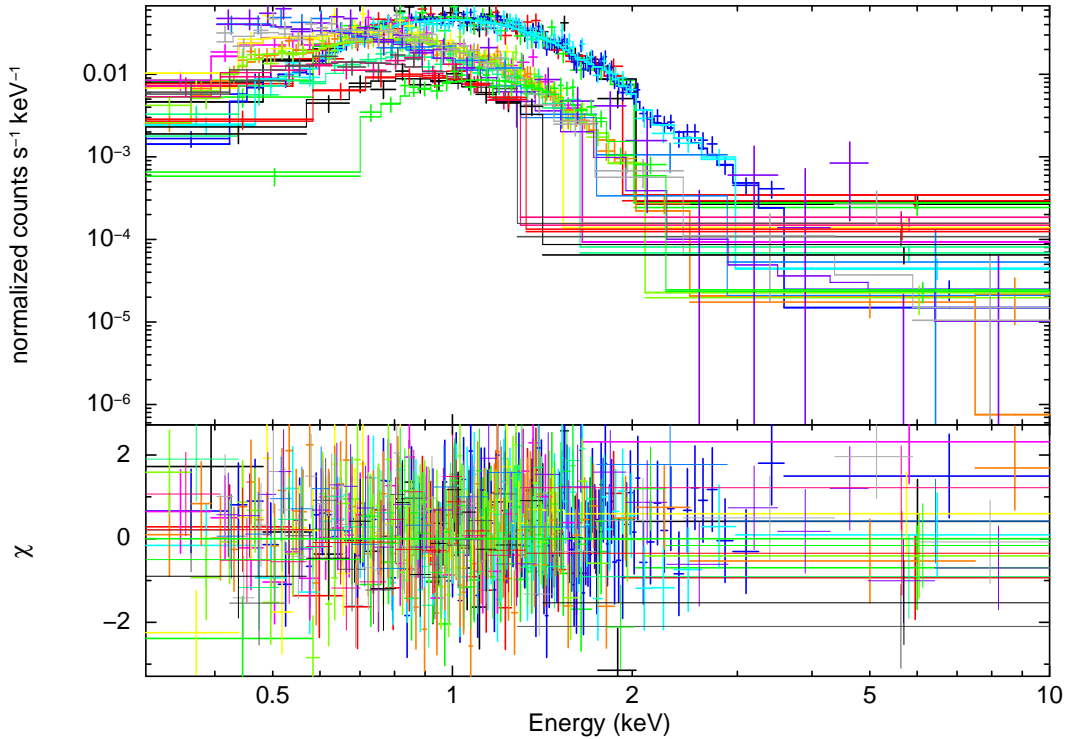}~}
  \bigskip
  \caption[]{\label{fig:Run7spectra} Figure showing the spectra
    resulting from Run \#7, obtained with the model {\tt
      wabs*(nsatmos+pow)}, for which the best-fit statistic is
    \Chisq{0.98}{628}{0.64}.}
\end{figure*}

In an effort to include all possible uncertainties in the production
of the \mr\ confidence regions, Gaussian Bayesian priors for the
source distances parameters are included, instead of keeping the
parameter values fixed\footnote{For a review on Bayesian analysis, see
  \cite{gregory05}}.  Since, additional systematic uncertainties can
arise when keeping the \nh\ parameters fixed, this assumption is also
relaxed in the spectral analysis.  Finally, the spectra of some qLMXBs
display excess flux above $2\keV$, which is not due to the
H-atmosphere thermal emission.  This is accounted for by adding a PL
spectral component to the model, where the photon index is fixed at
$\Gamma=1.0$ but the PL normalizations are free to vary.  Such PL
index is the hardest observed for a LMXB in quiescence
\citep[Cen~X-4][]{cackett10}.  The spectra resulting from this
analysis (with all five qLMXBs, and with all assumptions relaxed) are
shown in Figure~\ref{fig:Run7spectra}.

Relaxing the assumptions mentioned above adds 15 free parameters, for
a total of 27, which increases the complexity of the \chisq-space.
Because of that, in \emph{XSPEC}, the estimation of the confidence
region for each parameter proves difficult.  The command {\tt steppar}
iteratively calculates the \chisq-value for fixed values of a
parameter in the range provided by the user.  However, this
grid-search procedure is highly dependent on the starting point of the
parameter of interest and on the number of steps.  Such a problem is
particularly evident in the case of highly covariant sets of
parameters.  This can result in 1D or 2D \deltachisq\ contours that
are not reliable to estimate the uncertainties.  A solution to this
issue consists of using the posterior distributions from MCMC
simulations (described in \S~\ref{sec:mcmc}).  With those, one can
quantify the uncertainties of each parameters.  The need to
include Bayesian priors also brings forward the use of MCMC
simulations.

The simultaneous spectral fitting of all five targets using MCMC
simulations was performed in seven separate runs, during which the
assumptions on the spectral model are progressively relaxed.  The
characteristics of each run are described in Section~\ref{sec:radius}.
Another run is also performed with the {\tt nsagrav} model for
purposes of comparison with the {\tt nsatmos} model.  The following
subsection describes the MCMC analysis performed.

\subsection{Markov-Chain Monte Carlo Analysis}
\label{sec:mcmc}

As described above, the main advantage of using an MCMC simulation
resides in a complete understanding of the posterior probability
density functions of each parameter.  It also allows one to
marginalize over the so-called nuisance parameters, i.e., those that
are an important part of the modeling but which are of little physical
interest to the problem at hand.

Because of the curved parameter distributions obtained with the {\tt
  nsatmos} model, in particular the \mr\ contours, an MCMC algorithm
different from the typically used Metropolis-Hasting algorithm (MH) is
chosen. Indeed, we find that the MH algorithm is not efficient at
exploring skewed parameter spaces.  The next few paragraphs are
dedicated to a brief description of the Stretch-Move algorithm used.

The Stretch-Move algorithm \citep{goodman10} is particularly useful for
elongated and curved distributions (e.g., \mr\ with {\tt nsatmos}), as
demonstrated in previous works \citep[e.g.,][]{wang11}; our
implementation generally follows that work.  Other analyses have used
the Stretch-Move algorithm (\citealt{bovy12,olofsson12}, using another
implementation, by \citealt{foremanmackey12}).  The algorithm consists of
running several simultaneous chains, also called walkers, where the
next iteration for each chain is chosen along the line connecting the
current point of the chain and the current point of another randomly
selected chain.  The amount of ``stretching'' is defined by a random
number $z$ from an affine-invariant distribution \citep{goodman10}:
\begin{equation}
  g\left( z \right) \propto 
    \begin{cases}
      \frac{1}{\sqrt{z}} & \text{if } z \in \left[ \frac{1}{a},a\right]\\
      0 & \text{otherwise} \\
    \end{cases}
\end{equation}
In this work, each individual chain starts from a randomly selected
point in the parameter space, within the typical hard limits defined
for {\tt nsatmos} in \emph{XSPEC}.

The parameter $a$ is used as a scaling factor and is adjusted to
improve performance.  A larger value of $a$ increases the
``stretching'' of the chains, i.e., the algorithm will better explore
the elongated parts of the parameter space, but it decreases the
likelihood that the next step is accepted.  A smaller value of $a$
only produces small excursions from the previous value but increases
the likelihood that the next step is accepted.  Efficiency is
optimized at an intermediate value of $a$.  The Stretch-Move algorithm
can be fine-tuned with only two parameters: $a$ and the number of
simultaneous chain.  By comparison, the MH algorithm requires
$N(N+1)/2$ tuning variables, where N is the number of free parameters.

The validity of this MCMC algorithm is assessed by performed a test
run with a single source (U24 in NGC~6397, with fixed distance) and
comparing the resulting \mr\ contours with those obtained from a
simple grid-search method ({\tt steppar} in \emph{XSPEC}).
Specifically, the obtained \mr\ contours as well as other posterior
distributions match those obtained from a {\tt steppar} grid-search in
\emph{XSPEC}.  The addition of Gaussian Bayesian priors on the
distance is also tested with U24, which results in \mr\ contours
broadened in the \rinfty\ direction.  This is because the
normalization of the thermal spectrum is approximately $\propto \left(
\rinfty / d \right) ^{2}$.

For the Stretch-Move algorithm, the minimum number of simultaneous
chains is equal to N + 1, where N is the number of free parameters.
However, increasing the number of simultaneous chains ensures a more
complete coverage of the parameter space, when comparing the results
of the Stretch-Move algorithm to contours obtained with {\tt steppar}.
In addition, it reduces the chances of having the N + 1 walkers
collapsing to a N -- 1 dimensional space, i.e., one of the parameters
has the same value within all the chains causing all following steps
to evolve in the same plane.  However, increasing the number of
walkers also increases the convergence time.

\begin{deluxetable*}{cccccc}
  \tablecaption{\label{tab:GC} Globular Cluster Relevant Parameters}
  \tablewidth{0pt}
  \tabletypesize{\scriptsize}    
  \tablecolumns{6}
  \tablehead{  \colhead{Name} & \colhead{$d_{\rm GC}$ (kpc)} & \colhead{Method} & \nhtt\ (\xray) & \nhtt\ (HI) & \colhead{Reference} }
  \startdata
    M28      &  5.5\ppm0.3   & Horizontal Branch fitting & 0.256\ud{0.024}{0.024} & 0.24  & \cite{testa01} \\ 
    NGC~6397 &  2.02\ppm0.18 & Dynamical                 & 0.096\ud{0.017}{0.014} & 0.14  & \cite{rees96} \\
    M13      &  6.5\ppm0.6   & Dynamical                 & 0.008\ud{0.044}{0.007} & 0.011 & \cite{rees96} \\
    \OmCen   &  4.8\ppm0.3   & Dynamical                 & 0.182\ud{0.045}{0.042} & 0.09  & \cite{vandeven06} \\
    NGC~6304 &  6.22\ppm0.26 & Horizontal Branch fitting & 0.346\ud{0.105}{0.084} & 0.266 & \cite{recioblanco05} \\
  \enddata

  \tablecomments{The selection of the distance values is described in
    Section~\ref{sec:distances}, and the quoted uncertainties are
    $1\sigma$.  The \nh\ values are given in units of
    $\ee{22}\unit{atoms\percmsq}$, with 90\%-confidence uncertainties
    from \xray\ spectral fitting.  The \nh\ (HI) column corresponds to
    value in the direction of GCs, in the HI survey of
    \citep{dickey90}.  The \xray\ values are deduced from the best-fit
    \nh\ obtained from \xray\ spectral fitting of each target in this
    work.  Only the \nh\ values for NGC~6397 and \OmCen\ are not
    consistent with the HI values (see \S~\ref{sec:rinfty} for
    details).  \nh\ values deduced from the present \xray\ spectral
    analysis are used in the present work.}
\end{deluxetable*}

The resulting posterior distributions are then marginalized over
nuisance parameters.  While necessary for the spectral fitting, these
parameters do not provide physical information (e.g., $\alpha$, the
pile-up parameter).  The results are presented in Section 4, where the
values quoted correspond to the median value (i.e., 50\% quantile) of
each parameter.  The results are also presented in the
Figures~\ref{fig:M28contours}--\ref{fig:NGC6304contours} and
\ref{fig:fixNhfixDNoPL}--\ref{fig:nsagrav}, as one- and two-
dimensional posterior probability density distributions.  For each 1D
probability density distributions, we determine the 68\%, 90\%, and
99\% confidence regions using quantiles, which are delimited by the
solid, dashed and dotted lines in the 1D probability density
distribution of each figure
(Figures~\ref{fig:M28contours}--\ref{fig:NGC6304contours} and
\ref{fig:fixNhfixDNoPL}--\ref{fig:nsagrav}).  This ensures that the
integrated probabities on each side of the median are equal (i.e.,
equal areas under the probability density curves).  In addition, the
median value of some parameters are different from the most probable
value, especially in the case of highly skewed parameter posterior
distributions.  In some cases, the normalized probability of a
parameter posterior distribution does not converge to zero within the
parameter's hard limits in \emph{XSPEC}.  This is indicated by a '$p$'
in the tables listing the parameters.  The 2D posterior distributions
are normalized to unity and the color bars indicate the probability
density in each bin.  The 68\%, 90\%, and 99\% contours are obtained
by calculating the lines of constant probability density that enclose
68\%, 90\%, and 99\% of the accepted MCMC steps, respectively.

\subsection{Distances to the Globular Clusters and their Uncertainties}
\label{sec:distances}

While most GCs have distances estimated from photometry -- using RR
Lyrae variable stars \citep{marconi03,bono07}, horizontal branch
stars \citep{valenti07,gratton10}, or the carbon-oxygen white-dwarf
(CO-WD) sequence \citep{hansen07} -- these methods suffer from
systematic uncertainties that are difficult to quantify.  In fact,
many recent photometric studies of GCs do not quote the amount of
uncertainty in the measured distance
\citep{rosenberg00,bica06,gratton10}.

While some references discuss systematic uncertainties related to the
correction of extinction \citep[e.g.][for \OmCen]{mcdonald09}, other
sources of systematic errors can affect the results, including errors
related to the metallicity of cluster members (see dispersion in
Figure 1 of \citealt{harris10}), to a possible differential reddening
in the direction of GCs (as observed for \OmCen, \citealt{law03}), to
variations in the modeling of extinction with
$R\left(V\right)\sim3.1-3.6$ (adding $\sim10\%$ of uncertainty,
\citealt{grebel95}), or to the stellar evolution/atmosphere models
used.  As an example for the latter, distance determination methods to
NGC~6397 using CO-WD may be affected by uncertainties in the
evolutionary code models \citep{hansen07,strickler09}, which are not
easily quantifiable.

Therefore, whenever possible, dynamical distance measurements are used
-- distances estimated from proper motion and radial velocities of
cluster members.  These purely geometrical methods produce
well-understood uncertainties, although they are at the moment larger
than reported uncertainties from photometric methods.  This is
consistent with the goal of this paper which is to estimate \rns\ and
its uncertainties, minimizing systematic uncertainties.  The upcoming
mission GAIA from the European Space Agency, scheduled for 2013, is
expected to produce GC distance measurements, to an accuracy of few
percent, by determining the parallax of cluster members
\citep{baumgardt05,baumgardt08}.

\begin{deluxetable*}{clllllr}
  \tablecaption{\label{tab:rinfty} Spectral Fit Results of Individual Sources}
  \tablewidth{0pt}
  \tabletypesize{\scriptsize}    
  \tablecolumns{7}
  \tablehead{
    \colhead{Target} & \colhead{\kteff} & \colhead{\rns} & \colhead{\mns} &
    \colhead{\rinfty} & \nhtt &\colhead{$\chisqnu$/d.o.f. (prob.)}\\
    \colhead{} & \colhead{(keV)} & \colhead{(km)} & \colhead{(\Msun)} &
    \colhead{(km)} & \colhead{}
  }
  \startdata
    M28      & 120\ud{44}{12} & 10.5\ud{2.0}{2.9}  & 1.25\ud{0.54}{0.63p}  & 13.0\ud{2.3}{1.9} & 0.252\ud{0.025}{0.024}  & 0.94 / 269 (0.76) \\
    NGC~6397 &  76\ud{14}{7}  &  6.6\ud{1.2}{1.1p} & 0.84\ud{0.30}{0.28p}  &  8.4\ud{1.3}{1.1} & 0.096\ud{0.017}{0.015}  & 1.06 / 223 (0.25) \\
    M13      &  83\ud{26}{11} & 10.1\ud{3.7}{2.8p} & 1.27\ud{0.71}{0.63p}  & 12.8\ud{4.7}{2.4} & 0.008\ud{0.044}{0.007p} & 0.94 / 63  (0.62) \\
    \OmCen   &  64\ud{17}{7}  & 20.1\ud{7.4p}{7.2} & 1.78\ud{1.03p}{1.07p} & 23.6\ud{7.6}{7.1} & 0.182\ud{0.041}{0.047}  & 0.83 / 50  (0.80) \\
    NGC~6304 & 107\ud{32}{17} &  9.6\ud{4.9}{3.4p} & 1.16\ud{0.90}{0.56p}  & 12.2\ud{6.1}{3.8} & 0.346\ud{0.099}{0.093}  & 1.07 / 29  (0.36) \\
    \hline
    M28      & 119\ud{39}{9}  & 10.6\ud{0.9}{2.6}  & 1.17\ud{0.51}{0.56p} & 12.9\ud{0.9}{0.9} & (0.252) & 0.94 / 270 (0.77) \\ 
    NGC~6397 &  76\ud{15}{6}  &  6.6\ud{0.7}{1.1p} & 0.84\ud{0.24}{0.28p} &  8.4\ud{0.5}{0.5} & (0.096) & 1.06 / 224 (0.26) \\ 
    M13      &  86\ud{27}{10} &  9.2\ud{1.7}{2.3p} & 1.15\ud{0.42}{0.53p} & 11.6\ud{1.8}{1.5} & (0.008) & 0.93 /  64 (0.63) \\ 
    \OmCen   &  64\ud{13}{5}  & 19.6\ud{3.3}{3.8}  & 1.84\ud{0.98p}{1.10p}& 23.2\ud{3.6}{3.3} & (0.182) & 0.82 /  51 (0.82) \\ 
    NGC~6304 & 106\ud{31}{13} &  9.4\ud{2.4}{2.4p} & 1.12\ud{0.52}{0.51p} & 11.8\ud{2.5}{2.0} & (0.346) & 1.05 /  30 (0.39) \\ 
  \enddata

\tablecomments{ The targets were fit individually with fixed
  distances.  The top part shows the results of fits obtained with
  free values of \nh, while the bottom shows results obtained with
  fixed \nh\ (indicated in parenthesis).  For M28, the {\tt pileup}
  model is included (see \S~\ref{sec:analysis} for details), and a
  value $\alpha=0.45\ud{0.13}{0.13}$ is obtained.  The posterior
  distribution of \rinfty\ was obtained by calculating the value of
  \rinfty\ from \rns\ and \mns\ at each accepted MCMC iteration.
  Quoted uncertainties are 90\% confidence. ``p'' indicates that the
  posterior distribution did not converge to zero probability within
  the hard limits of the model.}
\end{deluxetable*}

The adopted distance values are discussed below and are summarized in
Table~\ref{tab:GC}.  In the following list, uncertainties are quoted
at the $1\sigma$ level (for GC distances, distance modulii, etc.)

\begin{itemize}
\item The GC M28 does not have a dynamical distance measurement, but
  its distance has been estimated in different works:
  5.1\ppm0.5\kpc\ \citep{rees91}, 4.8--5.0\kpc\ \citep{davidge96} and
  5.5\kpc\ \citep{harris96,testa01}, all using photometric methods.
  For the most recent result, uncertainties can be estimated from the
  uncertainties in the horizontal branch (HB) magnitude.
  Specifically, the uncertainty in $V_{\rm HB}=15.55\pm0.1$,
  translates into the uncertainty in the distance: $d_{\rm
    M28}=5.5\pm0.3\kpc$ \citep{servillat12}.  This measured value and
  its uncertainties were used here.

\item The distance to NGC~6397 has been reported from a dynamical
  study to be $d_{\rm NGC~6397}=2.02\ppm0.18\kpc$ \citep{rees96}.
  More recent photometric studies (CO WD sequence) have been
  performed, with $d=2.54\ppm0.07\kpc$ \citep{hansen07}, or
  $d=2.34\ppm0.13\kpc$ \citep{strickler09}, but since those results
  are model-dependent, they are not used in an effort to minimize
  unquantified systematics.  When the present analysis was at an
  advanced near-completion stage, recent results reporting a dynamical
  measurement of the distance came to our attention: $d_{\rm
    NGC~6397}=2.2\ud{0.5}{0.7}\kpc$ \citep{heyl12}, consistent with
  $d_{\rm NGC~6397}=2.02\ppm0.18\kpc$, the value used in the present
  work.

\item For M13, the dynamical distance has been measured: $d_{\rm M13}
  = 6.5\ppm0.6\kpc$ \citep{rees96}.  No other paper in the literature
  reports a distance measurement with quantified uncertainty.  This
  dynamical measurement is consistent with the value $d_{\rm M13} =
  7.1\kpc$ obtained form photometry \citep{harris96,sandquist10}.
  While uncertainties could be estimated for this measurement like it
  was done for M28, the dynamical measurement is preferred to limit the
  effect of systematic uncertainties, as explained above.

\item \OmCen's distance was measured in a dynamical study,
  $d_{\OmCen}=4.8\ppm0.3\kpc$ \citep{vandeven06}, and no other
  reference provides a distance with its measurement uncertainty.
  This measurement is consistent with other estimates
  (e.g. $d_{\OmCen}=5.2\kpc$, \citealt[][update 2010]{harris96}).

\item The GC NGC~6304 lacks a dynamical distance measurement.
  However, results from a previous work \citep[][using photometric
    data from \citealt{piotto02}]{recioblanco05} are available.  In
  that work, the distance modulus in the F555W filter
  (\hstlong\ filter) is $\left(m-M\right)_{F555W}=15.58\ppm0.09$.  The
  reddening in this band for NGC~6304 was not provided in the
  published work, but the value $E\left(B-V\right) = 0.52$
  \citep{piotto02} can be used instead.  This is acceptable because
  the average difference between $E\left({\rm HST}\right)$ and
  $E\left(B-V\right)$ in the \cite{recioblanco05} catalogue is $\Delta
  E=0.005$, which has a negligible effect on the absolute distance
  modulus.  Therefore, $\left(m-M\right)_{0}=13.97\ppm0.09$, assuming
  $A_{V}=3.1\times E\left(B-V\right)$, give $d_{\rm
    NGC~6304}=6.22\ppm0.26\kpc$.

\end{itemize}

Overall, the distances to the targeted GCs have uncertainties of
$\sim9\%$ or less, keeping in mind that the distances determined with
photometric methods possibly have systematically underestimated
uncertainties.

\section{Results}
\label{sec:results}

In this section, the results of the spectral analyses of each target
individually, with their \rinfty\ measurements, are first presented.
These include comparisons with previously published results.  In
particular, some issues regarding the reported spectral analyses for
the qLMXBs in \OmCen\ and M13 are raised.  Following this, the results
of the \rns\ measurement from the simultaneous fit are detailed.
 
\begin{figure*}[ht]
  \centerline{~\psfig{file=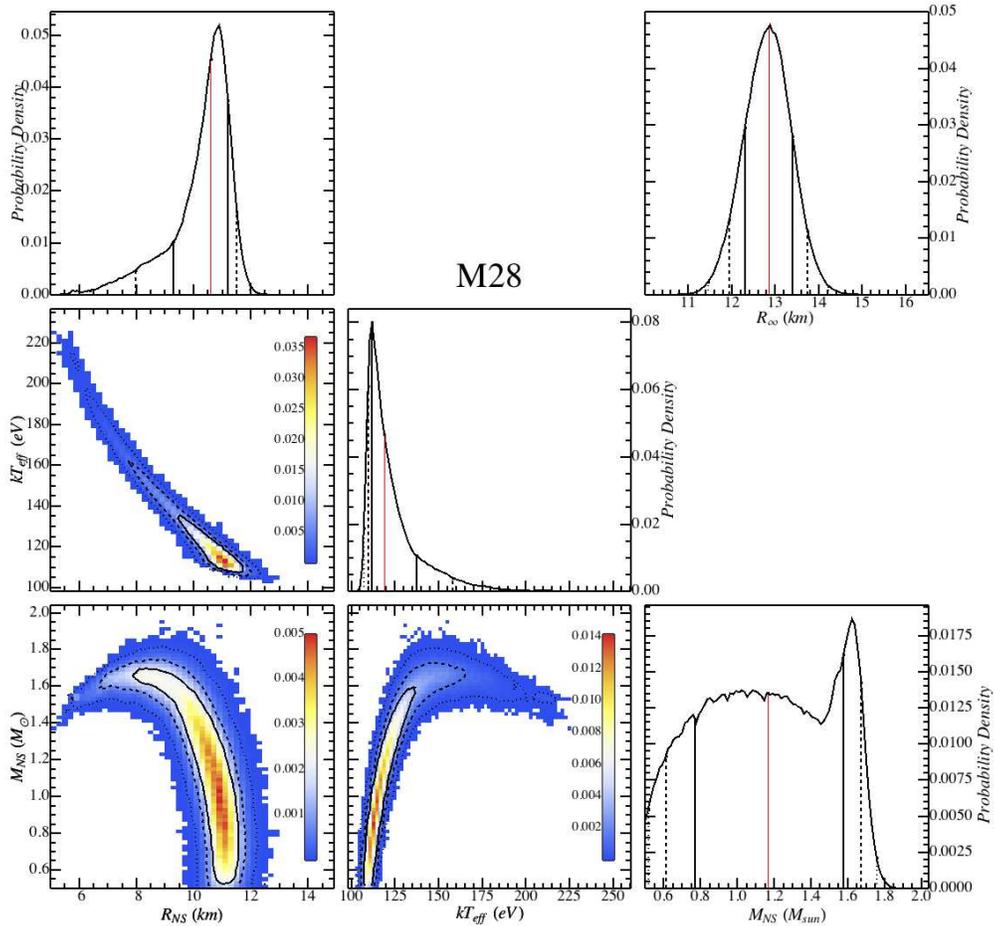,width=14cm,angle=0}~}
  \bigskip
  \caption[]{\label{fig:M28contours} Figure showing the one- and
    two- dimensional marginalized posterior distributions for the NS
    properties (radius, temperature and mass) obtained from the MCMC
    run for the qLMXB in M28, for fixed distance and \nh, i.e.,
    corresponding to the lower part of Table~\ref{tab:rinfty}.  The 1D
    and 2D posterior probability density distributions are normalized
    to unity.  The top-right plot shows the 1D posterior distribution
    of $\rinfty$ values.  The 68\%, 90\% and 99\% confidence intervals
    or regions are shown with solid, dashed, and dotted lines,
    respectively.  In the 1D distributions, the median value is shown
    as a red line. Note that the 99\% region is not always visible in
    the 1D distributions.  The physical radius of the NS in M28 is
    $\rns=10.6\ud{0.9}{2.6}\km$.  This corresponds to a projected
    radius of $\rinfty=12.9\ud{0.9}{0.9}\km$, for $\nhtt=0.252$.  The
    double-peaked 1D distribution of \mns\ is due to the strongly
    curved nature of the \mr\ and \mns--\kteff\ 2D distributions,
    i.e., the strong correlation between these parameters.  The color
    scale in each 2D distribution represents the probability density
    in each bin.  This figure and the following
    Figures~\ref{fig:NGC6397contours}--\ref{fig:nsagrav} were created
    with the Mathematica package LevelSchemes \citep{caprio05}.}
\end{figure*}

\subsection{\rinfty\ Measurements of Individual qLMXBs}
\label{sec:rinfty}
The analysis of the targeted qLMXBs is performed with the spectral
model detailed above (\S~\ref{sec:analysis}).  For each target,
analyzed individually, the fits are statistically acceptable (i.e.,
with a null hypothesis probability larger than 1\%), which
demonstrates that, within the statistics of the observations, the
sources did not experience any significant spectral variability over
the time scale between the observations.  The resulting values and
90\% confidence uncertainties, along with the \chisq-statistic
obtained, are provided in Table~\ref{tab:rinfty}.  The spectral
results obtained with \nh\ fixed at the \xray-deduced values, instead
of the usual HI survey values, are also provided.  Discrepancies
between the \xray-deduced and HI survey values of \nh, if any, are
discussed for each individual target.

Table~\ref{tab:rinfty} also shows the best-fit \rinfty\ values,
calculated using the equation:
\begin{equation}
\rinfty = \rns \left (1-\frac{2 G M_{\rm NS}}{\rns\ c^{2}},
\right)^{-1/2}
\end{equation}
from each accepted points of the MCMC runs.  Uncertainties in
\rinfty\ are then obtained from the calculated posterior distributions
of \rinfty\ resulting for the MCMC runs.  The use of MCMC simulations
has the advantage of avoiding geometrical construction to calculate
the uncertainties of \rinfty\ from the \mr\ contours as performed in
\cite{guillot11a}.

In the following subsections, the previously published results are
compared to those obtained here.  To do so, the \rinfty\ measurements
are renormalized to the distance used in the present analysis.

\begin{figure*}[ht]
  \centerline{~\psfig{file=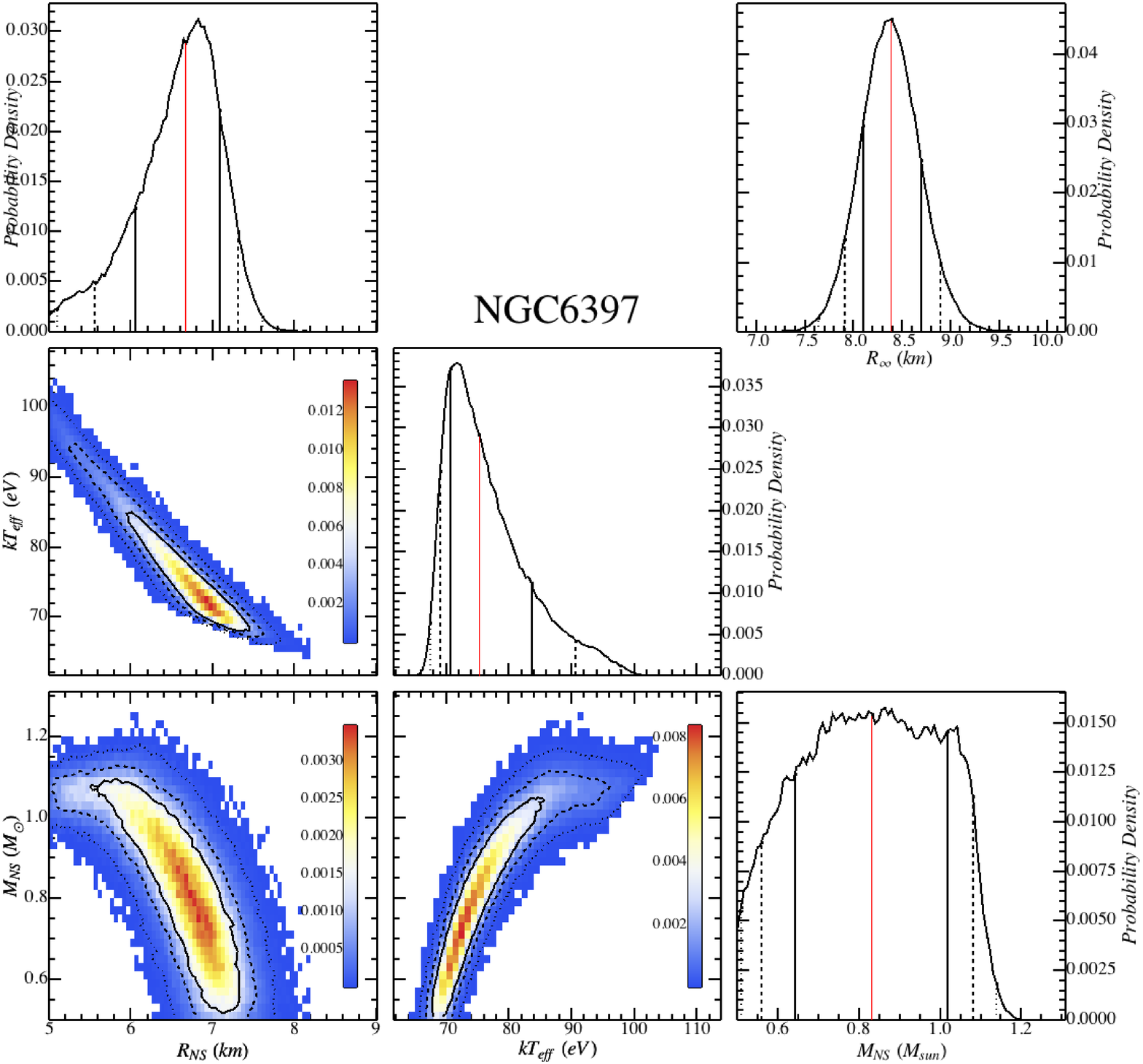,width=15cm,angle=0}~}
  \bigskip
  \caption[]{\label{fig:NGC6397contours} Figure similar to the
    precedent (Fig.~\ref{fig:M28contours}), but for the qLMXB in
    NGC~6397.  The physical radius of the NS is
    $\rns=6.6\ud{0.7}{1.1p}\km$ which corresponds to
    $\rinfty=8.4\ud{0.5}{0.5}\km$, for $\nhtt=0.096$.}
\end{figure*}
\begin{figure*}[ht]
  \centerline{~\psfig{file=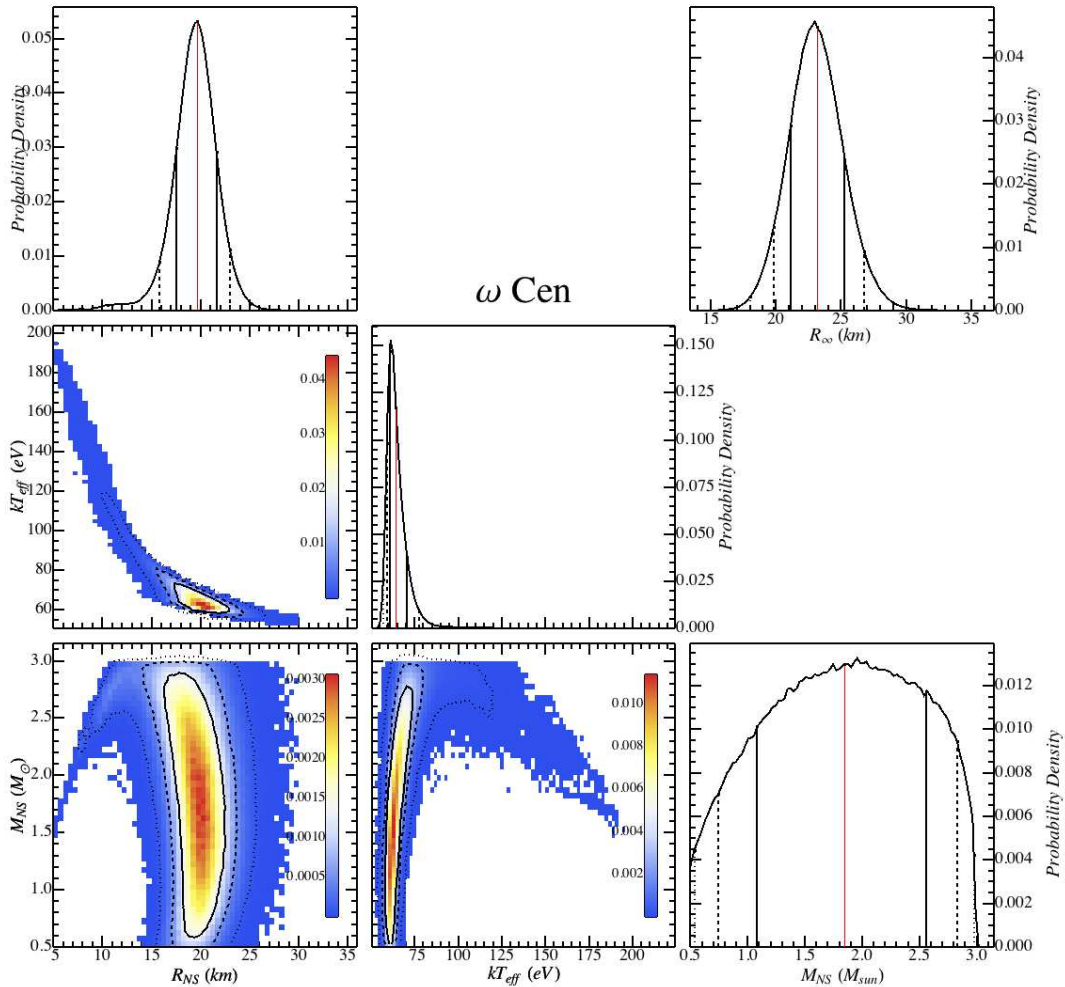,width=15cm,angle=0}~}
  \bigskip
  \caption[]{\label{fig:OmCencontours} Figure similar to
    Fig.~\ref{fig:M28contours}, but for the qLMXB in \OmCen.  The
    physical radius of the NS is $\rns=19.6\ud{3.3}{3.8}\km$ which
    corresponds to $\rinfty=23.2\ud{3.6}{3.3}\km$, for $\nhtt=0.182$.
  }
\end{figure*}
\begin{figure*}[ht]
  \centerline{~\psfig{file=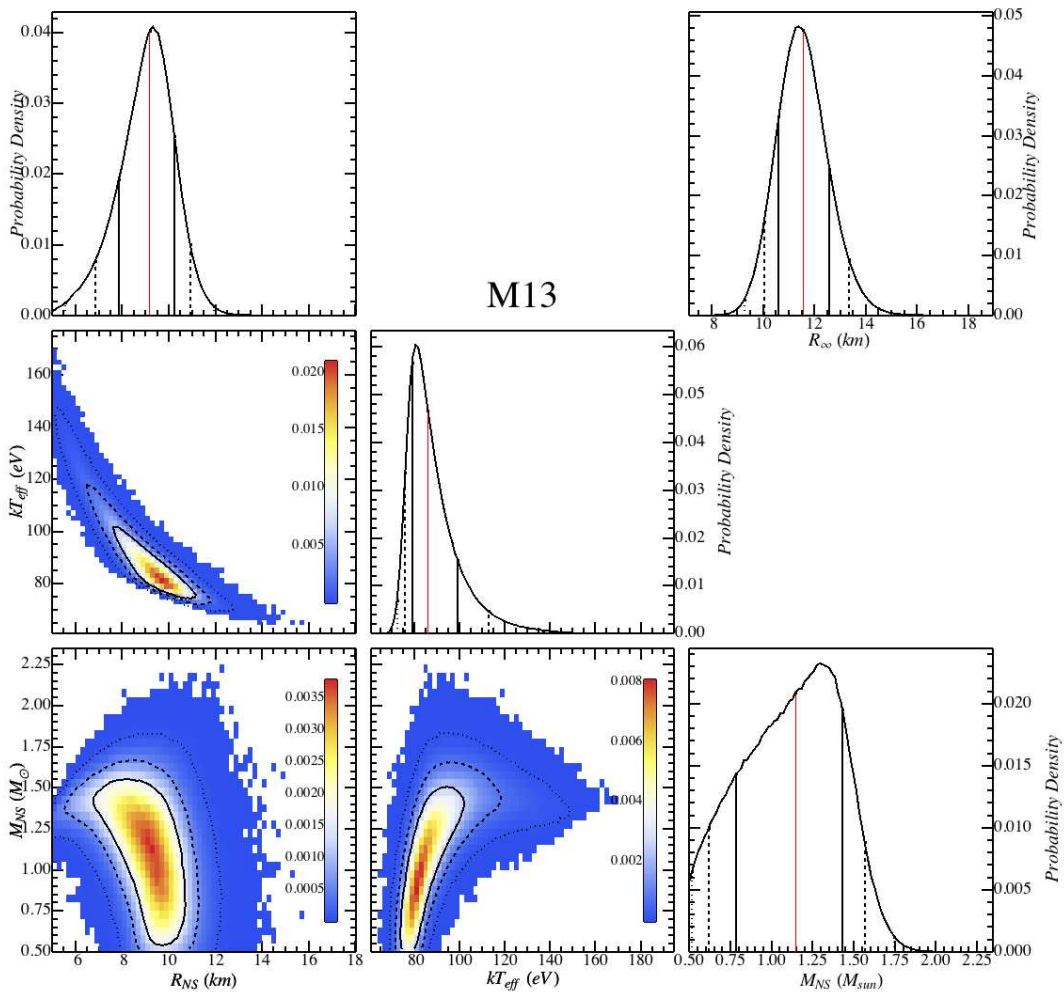,width=15cm,angle=0}~}
  \bigskip
  \caption[]{\label{fig:M13contours} Figure similar to
    Fig.~\ref{fig:M28contours}, but for the qLMXB in M13.  The
    physical radius of the NS is $\rns=9.2\ud{1.7}{2.3p}\km$ which
    corresponds to $\rinfty=11.6\ud{1.8}{1.5}\km$, for $\nhtt=0.008$.}
\end{figure*}
\begin{figure*}[ht]
  \centerline{~\psfig{file=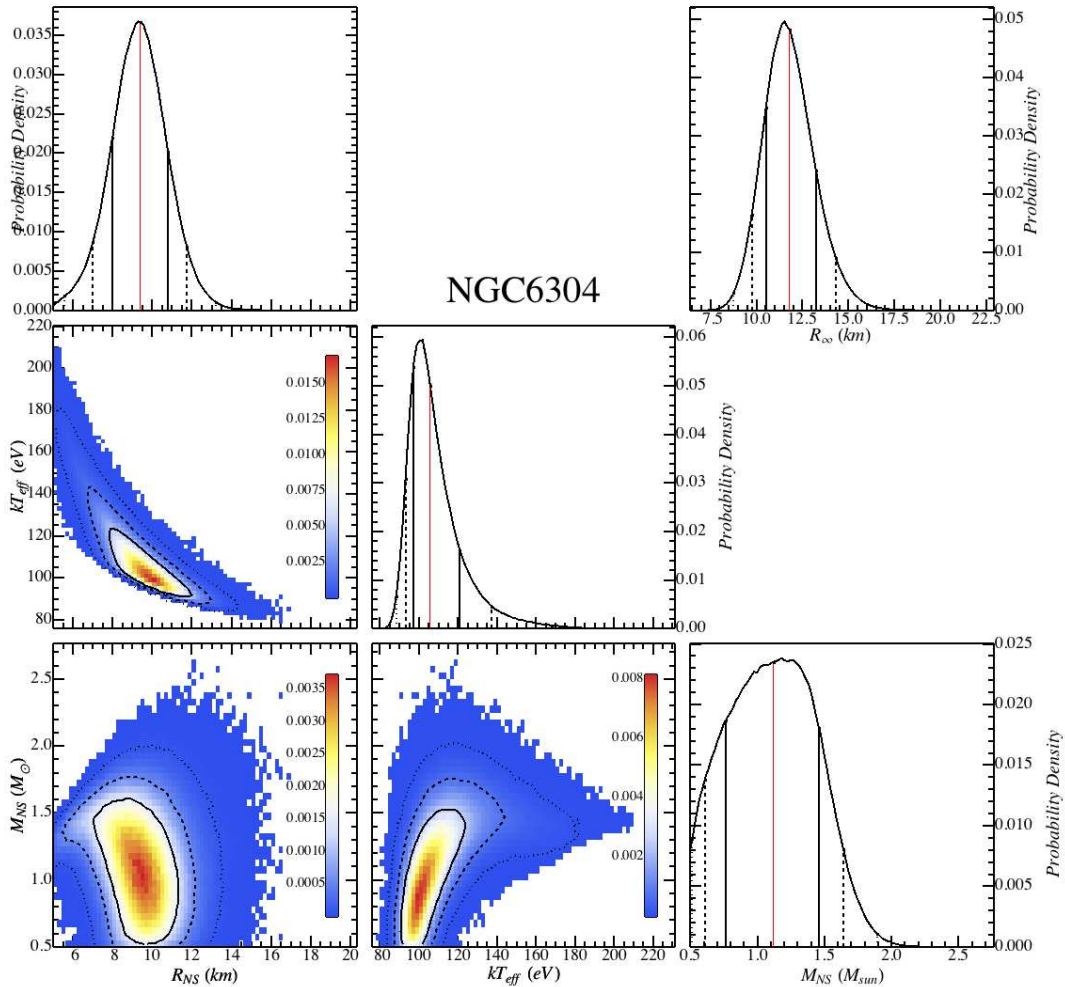,width=15cm,angle=0}~}
  \bigskip
  \caption[]{\label{fig:NGC6304contours} Figure similar to
    Fig.~\ref{fig:M28contours}, but for the qLMXB in NGC~6304.  The
    physical radius of the NS is $\rns=9.5\ud{2.4}{2.4p}\km$. The
    corresponding projected radius is $\rinfty=11.8\ud{2.5}{2.0}\km$,
    for $\nhtt=0.346$.}
\end{figure*}

\subsubsection{Comparison with Published Results - M28}
Using the 2002 \chandra\ data, the reported \rinfty\ value of the
qLMXB in M28 was
$\rinfty=14.5\ud{6.9}{3.8}\km\ \left(D/5.5\kpc\right)$
\citep{becker03}.  An additional 200\ksec\ of observations obtained
with \chandra-ACIS in 2008 was used to produce a refined radius
measurement: $\rns=9\ppm3\km$ and $\mns=1.4\ud{0.4}{0.9}\msun$, with
an H-atmosphere model \citep{servillat12}, corresponding to
$\rinfty=12.2\ud{2.6}{1.4}\km$ for $d_{\rm M28}=5.5\kpc$, consistent
with the discovery work \citep{becker03}.

All the NS parameters resulting from the present analysis
(Table~\ref{tab:rinfty},
$\rinfty=13.0\ud{2.3}{1.9}\km\ \left(D/5.5\kpc\right)$, for
$\nhtt=0.252\ud{0.025}{0.024}$) are also consistent with the
previously published results.  In addition, the previous work also
performed a careful variability analysis \citep{servillat12},
confirming our findings that the qLMXB in M28 is not variable.

The best-fit \nh\ found here is consistent with the value from an HI
survey: $\nhtt=0.24$ \citep{dickey90}, but the \xray-measured
\nh\ value is preferred in the rest of the present work, for the MCMC
runs with fixed \nh.

\subsubsection{Comparison with Published Results - NGC~6397}
The data sets used in this work are the same as the ones used in the
previous work \citep{guillot11a}.  There are however minor differences
in the data reduction, namely, the extraction radius used (99\% EEF in
this work compared to 98\% EEF at 1\keV\ previously), the calibration
files used (latest version of CALDB v4.4.8), the distance used for the
spectral fit, and the energy range (0.5--8\keV\ in
\citealt{guillot11a}).

After re-normalizing to the distance used in the present work, the
previous \rinfty\ result,
$\rinfty=9.6\ud{0.8}{0.6}\km\ \left(D/2.02\kpc\right)$, is consistent
with the one obtained from the MCMC run:
$\rinfty=8.4\ud{1.3}{1.1}\km\ \left(D/2.02\kpc\right)$, for
$\nhtt=0.096\ud{0.017}{0.015}$.  This best-fit value of \nh\ is
however, inconsistent with the fixed HI value ($\nh=0.14$,
\citealt{dickey90}) used in the previous work
\citep{guillot11a}\footnote{The \xray\ deduced value of \nh\ found
  here is nonetheless consistent with the \nh\ value from a different
  survey of Galactic HI \citep{kalberla05}, \nhtt=0.11, and with the
  \nh\ value calculated from the reddening in the direction of
  NGC~6397 \citep{harris96} with a linear relation between \nh\ and
  the extinction $A_{V}$ \citep{predehl95}.}.  This puts into question
the \rns\ measurement and \mr\ contours previously published with the
value \nhtt=0.14 \citep{guillot11a}.  When fixing \nhtt=0.14 in the
present work, the resulting \rinfty\ value is
$\rinfty=11.9\ud{0.8}{0.8}\km\ \left(D/2.02\kpc\right)$, marginally
consistent with the \citep{guillot11a} result.  Nonetheless, one
notices that the different value of \nh\ causes a significantly
different resulting \rns\ value.  Basically, increasing the assumed
value of \nh\ for a given target leads to a larger \rinfty.  This is
further discussed in Section~\ref{sec:discussion}.  In the rest of the
present work, the best-fit \xray\ deduced \nh\ value $\nhtt=0.096$ is
used.

\subsubsection{Comparison with Published Results - \OmCen}
The original \rinfty\ measurement from the \chandra\ discovery
observations was $\rinfty=14.3\ppm2.1\km\ \left(D/5.0\kpc\right)$ for
$\nhtt=0.09$ \citep{rutledge02b}, or
$\rinfty=13.7\ppm2.0\km\ \left(D/4.8\kpc\right)$.  Another work
measured $\rinfty=13.6\ppm0.3\km\ \left(D/5.3\kpc\right)$ with
$\nhtt=0.09\pm0.025$, equivalent to
$\rinfty=12.3\ppm0.3\km\ \left(D/4.8\kpc\right)$, using the
\xmm\ observation of \OmCen\ \citep{gendre03a}.  Results from both
analyses are consistent with the radius measurement performed in this
work, with the value of Galactic absorption \nhtt=0.09:
$\rinfty=11.9\ud{1.6}{1.4}\km\ \left(D/4.8\kpc\right)$.  However, when
removing the constraint on \nh, the best-fit \rinfty\ and \nh\ become
inconsistent with the previously reported values.  Specifically,
$\rinfty=23.6\ud{7.6}{7.1}\km\ \left(D/4.8\kpc\right)$ for
$\nhtt=0.182\ud{0.041}{0.047}$. This value of \nh, not consistent with
the HI survey value \citep{dickey90}, was used in the remainder of the
present work.  One can also note that the present results (best-fit
\rns, \mns, \kteff, and \nh) are consistent with those previously
published \citep{webb07}.

The results presented in Table~\ref{tab:rinfty} should be treated as
more realistic than the initially reported one since they make use of
more recent calibrations of \xmm\ and \chandra, as well as an improved
method.  In particular, the small uncertainties ($\sim2\%$) on
\rinfty\ previously published \citep{gendre03a} are particularly
intriguing. It has also been shown in another reference that the S/N
obtained with 50\ksec\ exposure of \OmCen\ is not sufficient to
constrain the radius with $\sim 2\%$ uncertainty \citep{webb07}, but
the cause of this discrepancy was not discussed.  The constrained
\rns\ measurement with $\sim 2\%$ uncertainties \citep{gendre03a} was
not reproduced in the later work \citep[][$\rns=11.7\ud{7.0}{5.0}\km$,
  using the same \xmm\ data]{webb07}, nor in the present work.  Using
the same model as the one initially used \citep{gendre03a}, similar
uncertainties ($\sim2\%$) can only be obtained when keeping the NS
surface temperature fixed, leaving the normalization (i.e., the
projected radius \rinfty) as the sole free parameter.  Specifically,
with the same model and analysis procedure, the uncertainties on
\rinfty\ are $\sigma_{\rinfty}\sim3\%$ with the temperature fixed and
becomes $\sigma_{\rinfty}\sim15\%$ when the temperature is a free
parameter.  If this is the method used in \cite{gendre03a}, the
uncertainties of \rinfty\ only represent the statistical uncertainties
and are therefore highly underestimated.  It is inappropriate to keep
the temperature fixed because there is no known prior on the NS
surface temperature, and therefore it must remain free during the
spectral fitting.  In addition, the \xmm-pn observations suffer from
periods of high-background activity which need to be removed (see
Figure~\ref{fig:LC_OmCen}).  This leads to 34\ksec\ of usable exposure
time of the 41\ksec\ available.  No such background flares were
reported in the original works \citep{gendre03a,webb07}.

This note about the amount of uncertainty for \OmCen\ is of crucial
importance since this source has often been cited as the canonical
qLMXB, with the best radius measurement available, citing the
underestimated $\sim2\%$ uncertainties on \rns.  Deeper exposures of
\OmCen\ are needed to provide constraints that will be useful for dEoS
determination.  Moreover, this discussion also points out the
importance of reporting \mr\ contours (instead of simple
\rns\ measurements) for the measurements of NS properties using the
thermal emission from qLMXBs.

\begin{deluxetable*}{ccccccc}
  \tablecaption{\label{tab:SimulFixNh} Results from Simultaneous Spectral Fitting, with Fixed \nh}
  \tablewidth{0pt}
  \tabletypesize{\scriptsize}    
  \tablecolumns{7}
  \tablehead{
    \colhead{Target} & \colhead{$\alpha_{\rm pileup}$} & 
    \colhead{\kteff} & \colhead{\mns} & \colhead{\rinfty} & \colhead{\nhtt} & \colhead{PL Norm $\tee{-7}$}\\
    \colhead{} & \colhead{} & \colhead{(eV)} & \colhead{(\Msun)} & 
    \colhead{(\km)} & \colhead{} & \colhead{$\PLunit$}\\}
  \startdata
   \multicolumn{7}{c}{Run~\#1: {\tt Fixed \nh, Fixed $d_{\rm GC}$, No PL included}, \rns=7.1\ud{0.5}{0.6}\km}\\
   \multicolumn{7}{c}{\Chisq{0.97}{643}{0.70}, 18\% accept. rate} \\  
   \hline
   M28      & 0.44\ud{0.11}{0.11} & 176\ud{14}{11} & 1.62\ud{0.08}{0.08}  & 12.5\ud{0.6}{0.6} & (0.252) & -- \\ 
   NGC~6397 &       --            &  71\ud{7}{3}   & 0.69\ud{0.26}{0.16p} &  8.4\ud{0.5}{0.5} & (0.096) & -- \\ 
   M13      &       --            & 110\ud{12}{10} & 1.41\ud{0.21}{0.29}  & 11.0\ud{1.4}{1.3} & (0.008) & -- \\ 
   \OmCen   &       --            & 164\ud{14}{14} & 2.05\ud{0.13}{0.15}  & 18.9\ud{1.7}{1.7} & (0.182) & -- \\ 
   NGC~6304 &       --            & 136\ud{18}{17} & 1.41\ud{0.25}{0.43p} & 11.0\ud{1.8}{1.8} & (0.346) & -- \\ 
   \hline
   \hline
   \multicolumn{7}{c}{Run~\#2: {\tt Fixed \nh, Gaussian Bayesian priors for $d_{\rm GC}$, No PL included}, \rns=7.6\ud{0.9}{0.9}\km}\\
   \multicolumn{7}{c}{\Chisq{0.98}{638}{0.64}, 11\% accept. rate} \\  
   \hline
   M28      & 0.44\ud{0.11}{0.10} & 165\ud{22}{20} & 1.63\ud{0.14}{0.15}  & 12.6\ud{1.}{1.0} & (0.252) & -- \\ 
   NGC~6397 &       --            &  71\ud{8}{3}   & 0.73\ud{0.33}{0.20p} &  9.0\ud{1.1}{0.9} & (0.096) & -- \\ 
   M13      &       --            & 101\ud{20}{16} & 1.34\ud{0.33}{0.53p} & 11.0\ud{1.9}{1.7} & (0.008) & -- \\ 
   \OmCen   &       --            & 154\ud{21}{22} & 2.16\ud{0.22}{0.21}  & 19.3\ud{2.3}{2.1} & (0.182) & -- \\ 
   NGC~6304 &       --            & 127\ud{23}{19} & 1.36\ud{0.34}{0.59p} & 11.0\ud{2.1}{1.8} & (0.346) & -- \\ 
   \hline
   \hline	
   \multicolumn{7}{c}{Run~\#3: {\tt Fixed \nh, Fixed $d_{\rm GC}$, PL included}, \rns=7.3\ud{0.5}{0.6}\km}\\
   \multicolumn{7}{c}{\Chisq{0.96}{638}{0.78}, 15\% accept. rate} \\
   \hline
   M28      & 0.35\ud{0.12}{0.12} & 170\ud{14}{11} & 1.63\ud{0.08}{0.08}  & 12.6\ud{0.6}{0.6} & (0.252) & 5.1\ud{3.7}{3.4p} \\ 
   NGC~6397 &       --            &  70\ud{7}{3}   & 0.68\ud{0.28}{0.15p} &  8.6\ud{0.5}{0.5} & (0.096) & 2.2\ud{1.3}{1.3p} \\ 
   M13      &       --            & 109\ud{12}{11} & 1.50\ud{0.24}{0.32}  & 11.7\ud{1.8}{1.5} & (0.008) & 2.2\ud{4.2}{2.0p} \\ 
   \OmCen   &       --            & 163\ud{13}{14} & 2.14\ud{0.14}{0.16}  & 19.9\ud{1.9}{1.9} & (0.182) & 2.9\ud{3.7}{2.4p} \\ 
   NGC~6304 &       --            & 136\ud{18}{17} & 1.51\ud{0.26}{0.44p} & 11.7\ud{2.0}{2.0} & (0.346) & 1.7\ud{2.5}{1.5p} \\ 
   \hline
   \hline
   \multicolumn{7}{c}{Run~\#4: {\tt Fixed \nh, Gaussian Bayesian priors for $d_{\rm GC}$, PL included}, \rns=8.0\ud{1.0}{1.0}\km}\\
   \multicolumn{7}{c}{\Chisq{0.97}{633}{0.72}, 11\% accept. rate} \\ 
   \hline
   M28      & 0.35\ud{0.12}{0.12} & 157\ud{24}{20} & 1.64\ud{0.15}{0.18}  & 12.8\ud{1.0}{1.0} & (0.252) & 5.0\ud{3.8}{3.4p} \\ 
   NGC~6397 &       --            &  70\ud{8}{3}   & 0.72\ud{0.37}{0.19p} &  9.4\ud{1.1}{1.0} & (0.096) & 2.2\ud{1.3}{1.3p} \\ 
   M13      &       --            &  99\ud{22}{17} & 1.43\ud{0.37}{0.61p} & 11.7\ud{2.3}{1.9} & (0.008) & 2.2\ud{4.0}{1.9p} \\ 
   \OmCen   &       --            & 151\ud{21}{21} & 2.28\ud{0.25}{0.25}  & 20.4\ud{2.6}{2.4} & (0.182) & 3.0\ud{3.8}{2.5p} \\ 
   NGC~6304 &       --            & 125\ud{24}{20} & 1.46\ud{0.36}{0.69p} & 11.8\ud{2.2}{2.1} & (0.346) & 1.7\ud{2.4}{1.5p} \\ 
  \enddata

  \tablecomments{ $\alpha_{\rm pileup}$ corresponds to the parameter
    of the {\tt pileup} model.  ``PL Norm.'' refers to the value of
    the normalization of the power-law component, when used.  For each
    run, the characteristics are described: whether or not the
    absorption \nh\ was fixed; whether the GC distances $d_{\rm GC}$
    were fixed or if a Bayesian prior was imposed; whether or not a
    additional power-law component (PL) was included in the model.
    For each run, the best \chisqnu\ value is provided, as well as the
    null hypothesis probability.  Finally, the acceptance rate (not
    including the burn-in period) is provided.  All quoted
    uncertainties are 90\% confidence.  Values in parentheses are kept
    fixed in the analysis. ``p'' indicates that the posterior
    distribution did not converge to zero probability within the hard
    limit of the model.}
\end{deluxetable*}

\subsubsection{Comparison with Published Results - M13}
The \rinfty\ value of the qLMXB in M13 reported in the discovery paper
\citep{gendre03b}, $\rinfty=12.8\pm0.4\km\ \left(D/7.7\kpc\right)$,
corresponds to $\rinfty=10.8\pm0.3\km\ \left(D/6.5\kpc\right)$.  This
value is consistent with the value presented in the present work,
given the uncertainties:
$\rinfty=12.8\ud{4.7}{2.4}\km\ \left(D/6.5\kpc\right)$, for
$\nhtt=0.008\ud{0.044}{0.007p}$.  The best fit \nh\ is consistent with
HI survey values ($\nh=0.011$, \citealt{dickey90}), but $\nhtt=0.008$
is used in the remainder of the present analysis.

Once again, the uncertainties reported in the original work are small
and can only be reproduced when fixing the temperature.  Similarly to
\OmCen, the M13 discovery analysis was likely performed keeping the
temperature frozen to estimate the uncertainty on the radius, and the
$\pm0.3\km$ uncertainties cited \citep{gendre03b} are only systematic
uncertainties.

In summary, the results found in this work for M13 are consistent with
the existing ones \citep{gendre03b,webb07}, and while our radius
measurement uncertainties are not as constraining as those previously
reported, they are considered more realistic given the S/N available
for the observations, and given that more recent calibrations have
been used.  Similarly to \OmCen, deeper exposures of M13 would provide
the necessary S/N to constrain the dEoS.

When the present work was at an advanced stage, results of a spectral
analysis of the qLMXB in M13 came to our attention
\citep{catuneanu13}.  These results are consistent with those found in
the present work, when re-normalized to the distance used here.
 
\begin{figure*}[ht]
  \centerline{~\psfig{file=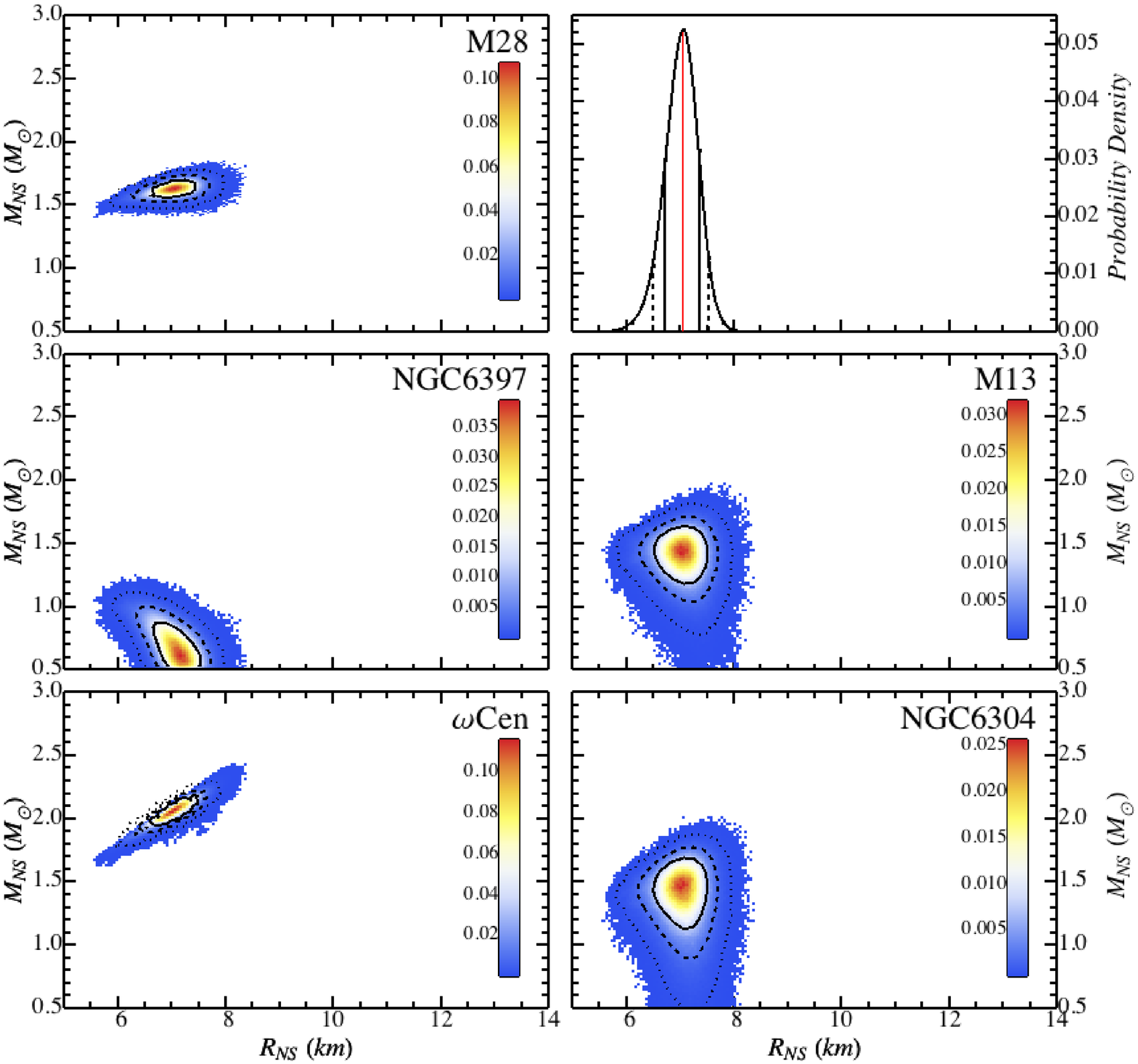,width=16cm,angle=0}~} 
  \caption[]{\label{fig:fixNhfixDNoPL} Figure showing the marginalized
    posterior distribution in \mr\ space for the five qLMXBs, in the
    first MCMC run, where the distance and the hydrogen column density
    \nh\ are fixed and where no PL component is added, corresponding
    to Run~\#1.  The 1D and 2D posterior probability distributions are
    normalized to unity.  The color scale in the 2D distributions
    represents the probability density in each bin.  The 68\%, 90\%
    and 99\%-confidence contours are shown with solid, dashed and
    dotted lines on the \mr\ density plots, respectively.  The
    top-right graph is the resulting normalized probability
    distribution of \rns, common to the five qLMXBs, with the 68\%,
    90\% and 99\%-confidence regions represented by the solid, dashed
    and dotted vertical lines.  The median value is shown by the red
    line.  The measured radius is $\rns=7.1\ud{0.5}{0.6}\km$ (90\%
    confidence).}
\end{figure*}

\subsubsection{Comparison with published results - NGC~6304}
This analysis presents a new 100\ksec\ observation of NGC~6304.  The
\rinfty\ value of the qLMXB,
$\rinfty=12.2\ud{6.1}{3.8}\km\ \left(D/6.22\kpc\right)$, for
$\nhtt=0.346\ud{0.099}{0.093}$), is consistent with that obtained from
the \xmm\ observation \citep{guillot09a},
$\rinfty=12.1\ud{6.6}{4.8}\km\ \left(D/6.22\kpc\right)$, and with that
from a short \chandra\ observation \citep{guillot09b},
$\rinfty=7.8\ud{8.6}{3.8}\km\ \left(D/6.22\kpc\right)$, after
re-normalizing the 2009 measurement to the distance used in the
present paper.  The best-fit \xray\ deduced value is also consistent
with the value used in the original work, obtained from HI surveys.
Nonetheless, the \xray\ measured \nh\ value is used in the remainder of
the present work.

\begin{figure*}[ht]
  \centerline{~\psfig{file=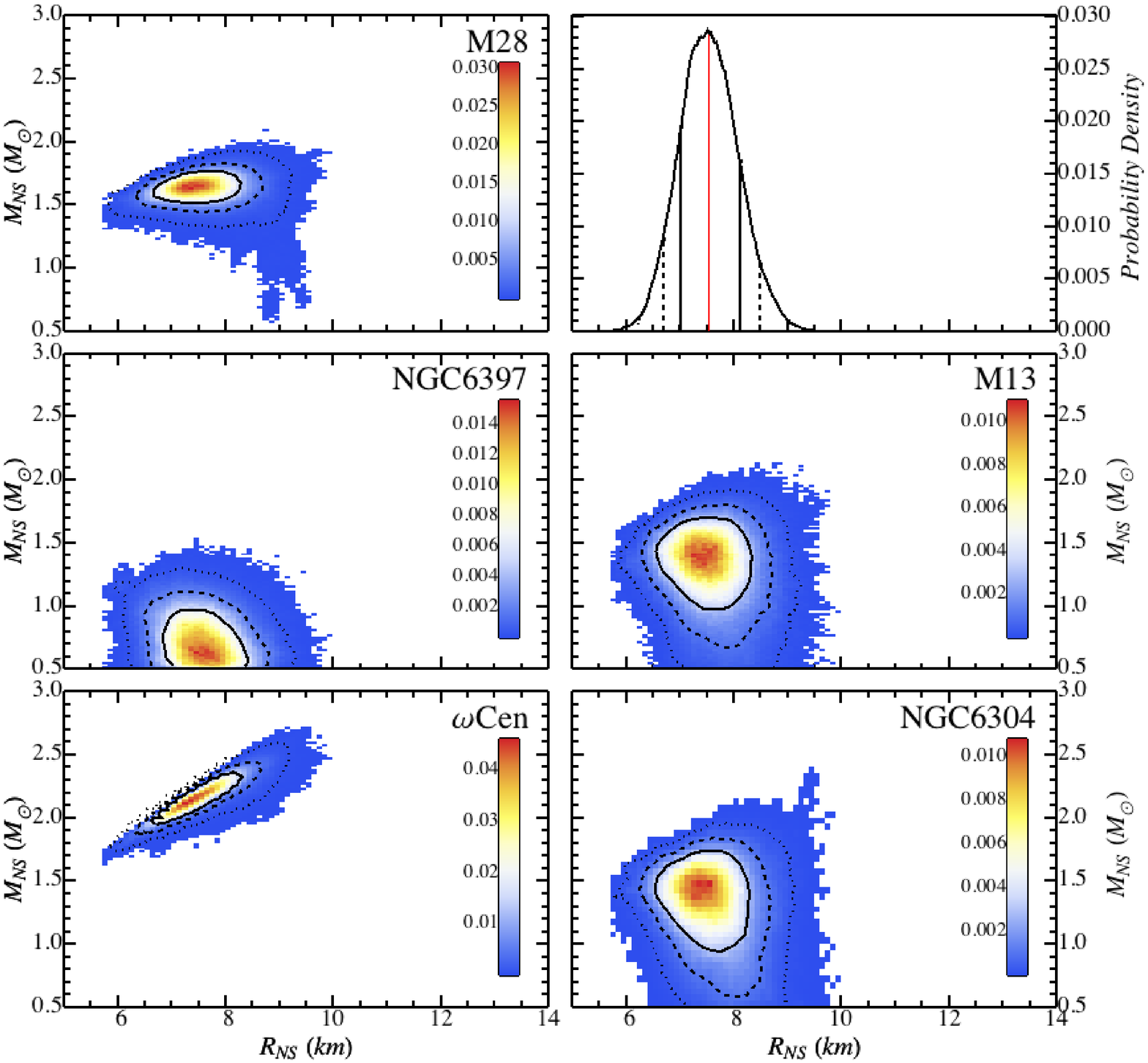,width=16cm,angle=0}~}
  \caption[]{\label{fig:fixNhFreeDNoPL} Figure similar to the previous
    one, Fig~\ref{fig:fixNhfixDNoPL}, but for the MCMC Run \#2, where
    Gaussian Bayesian priors were used for the distances to the five
    qLMXBs (see Table~\ref{tab:GC}). The resulting radius measurement
    is $\rns=7.6\ud{0.9}{0.9}\km$.}
\end{figure*}
\begin{figure*}[ht]
  \centerline{~\psfig{file=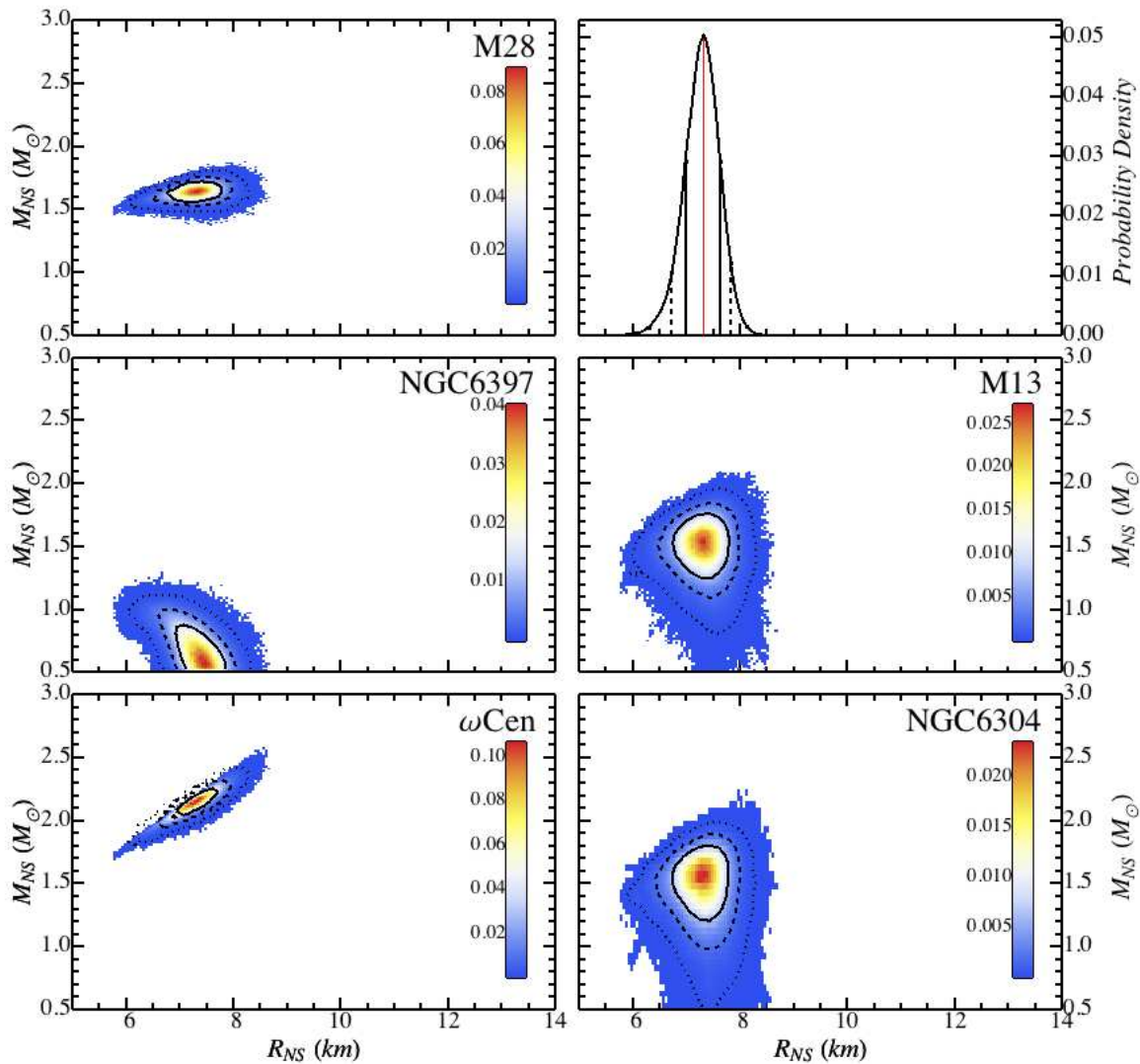,width=16cm,angle=0}~} 
  \caption[]{\label{fig:FixNhFixDwithPL} Figure similar to
    Figure~\ref{fig:fixNhfixDNoPL}, corresponding to the results of
    Run \#3.  In this run, the distances are fixed (no priors
    included), but a PL component (with fixed index $\Gamma=1$) is
    added to the spectral model, leading to
    $\rns=7.3\ud{0.5}{0.6}\km$.}
\end{figure*}
\begin{figure*}[ht]
  \centerline{~\psfig{file=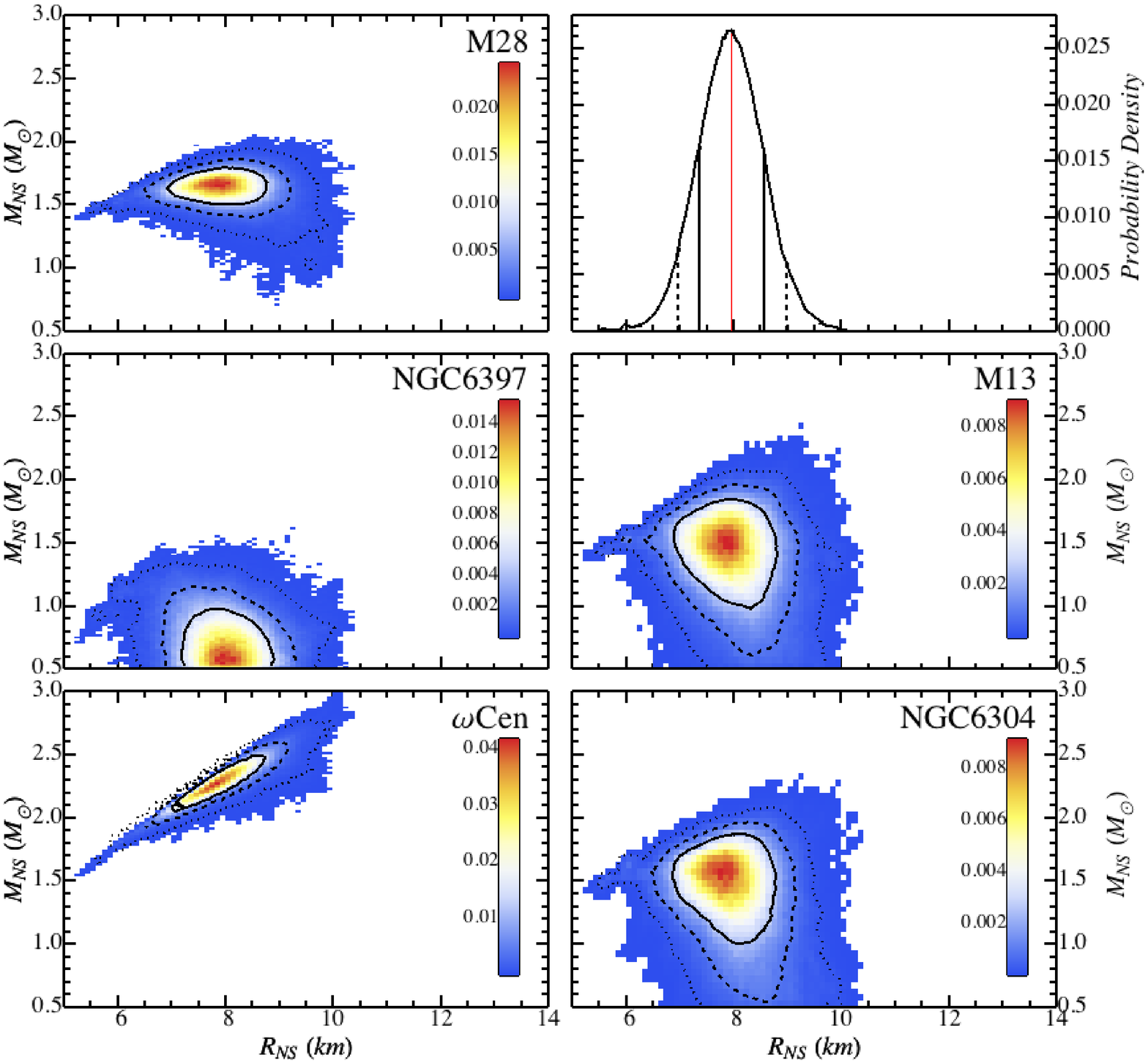,width=16cm,angle=0}~} 
  \caption[]{\label{fig:FixNhFreeDwithPL} Figure similar to
    Figure~\ref{fig:fixNhfixDNoPL}, but for Run \#4.  Here, the
    Gaussian Bayesian priors on the distances are included, as well as the PL
    spectral component, with \nh\ held fixed.  The resulting NS radius
    is $\rns=8.0\ud{1.0}{1.0}\km$.}
\end{figure*}

\subsection{\rns, Measurement of the Radius of Neutron Stars}
\label{sec:radius}
In this section, the results of the simultaneous spectral fits with
the parameter posterior distributions obtained from the MCMC
simulations are presented.  The following distinct MCMC runs are
performed:
\begin{itemize}
\item Run~\#1: Model {\tt nsatmos} with fixed \nh\ values and fixed
  distances: 12 free parameters, 25 Stretch-Move walkers.
\item Run~\#2: Model {\tt nsatmos} with fixed \nh\ values and Gaussian
  Bayesian  priors for the distances: 17 free parameters, 30
  Stretch-Move walkers.
\item Run~\#3: Model {\tt nsatmos} with fixed \nh\ values and fixed
  distances, and an additional PL component: 17 free parameters, 30
  Stretch-Move walkers.
\item Run~\#4: Model {\tt nsatmos} with fixed \nh\ values, Gaussian Bayesian 
  priors for the distances, and an additional PL component (with fixed
  index $\Gamma=1.0$, but free normalizations): 22 free parameters, 35
  Stretch-Move walkers.
\item Run~\#5: Model {\tt nsatmos} with free \nh\ values and fixed distances:
  17 free parameters, 30 Stretch-Move walkers.
\item Run~\#6: Model {\tt nsatmos} with free \nh\ values and Gausssian
  Bayesian priors for the distances: 22 free parameters,
  35 Stretch-Move walkers.
\item Run~\#7: Model {\tt nsatmos} with free \nh\ values, Gaussian
  Bayesian priors for the distances, and an additional PL component:
  27 free parameters, 40 Stretch-Move walkers. The spectra resulting
  from this run are shown in Figure~\ref{fig:Run7spectra}.
\item Run~\#8: Model {\tt nsagrav} with fixed \nh\ values and fixed
  distances: 12 free parameters, 25 Stretch-Move walkers.  This model
  is used for comparison with the {\tt nsatmos} model.
\end{itemize}
All the runs converged to a statistically acceptable point in the
parameter space, with $\chisqnu\sim1$ and a null hypothesis
probability $>0.01$.  In addition, the acceptance rate of each run is
large enough ($>5\%$) that the model used and the assumptions are
adequate for the data.  From the accepted steps of each run, the
marginalized posterior distributions of all parameters and the median
values with 90\% confidence regions are quoted in the tables.  In
addition, \mr\ posterior distributions for each of the five qLMXBs are
obtained.

Run~\#1 was performed with the maximum constraints imposed on the
model.  With the assumptions imposed on the parameters, this run leads
to the most constrained \mr\ contours of this work, resulting in
$\rns=7.1\ud{0.5}{0.6}\km$ (90\%-confidence), and the minimum
\Chisq{0.97}{643}{0.70}.  Detailed information for other parameters is
shown in Table~\ref{tab:SimulFixNh}.

The possible effect of auto-correlation between the steps of the MCMC
simulation is investigated by selecting every other 10 accepted point,
a method called thinning \citep{maceachern94}.  The resulting
confidence regions for all parameters are not affected by thinning and
it can be safely concluded that the steps in the MCMC runs are not
subject to auto-correlation.  All the accepted steps of each run are
therefore used to create the posterior distributions.

\begin{figure*}[ht]
  \centerline{~\psfig{file=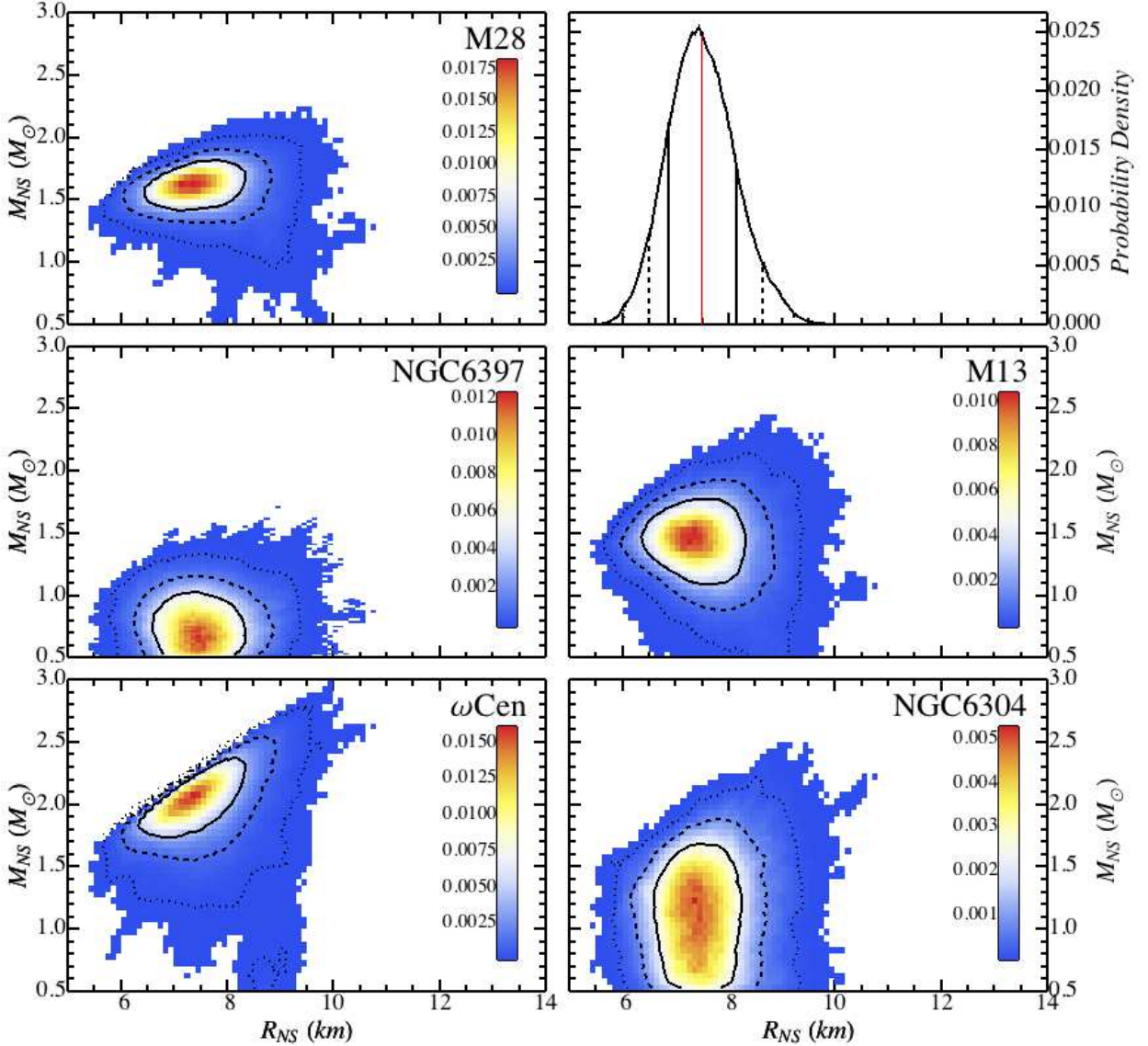,width=17cm,angle=0}~}
  \caption[]{\label{fig:FreeNhFixDNoPL} Figure similar to
    Figure~\ref{fig:fixNhfixDNoPL}, showing results of Run \#5.  The
    hydrogen column density \nh\ is left free in this run, but the
    distances remained fixed, and no PL component was added.  This run
    produced $\rns=7.5\ud{1.1}{1.0}\km$.}
\end{figure*}

\begin{figure*}[ht]
  \centerline{~\psfig{file=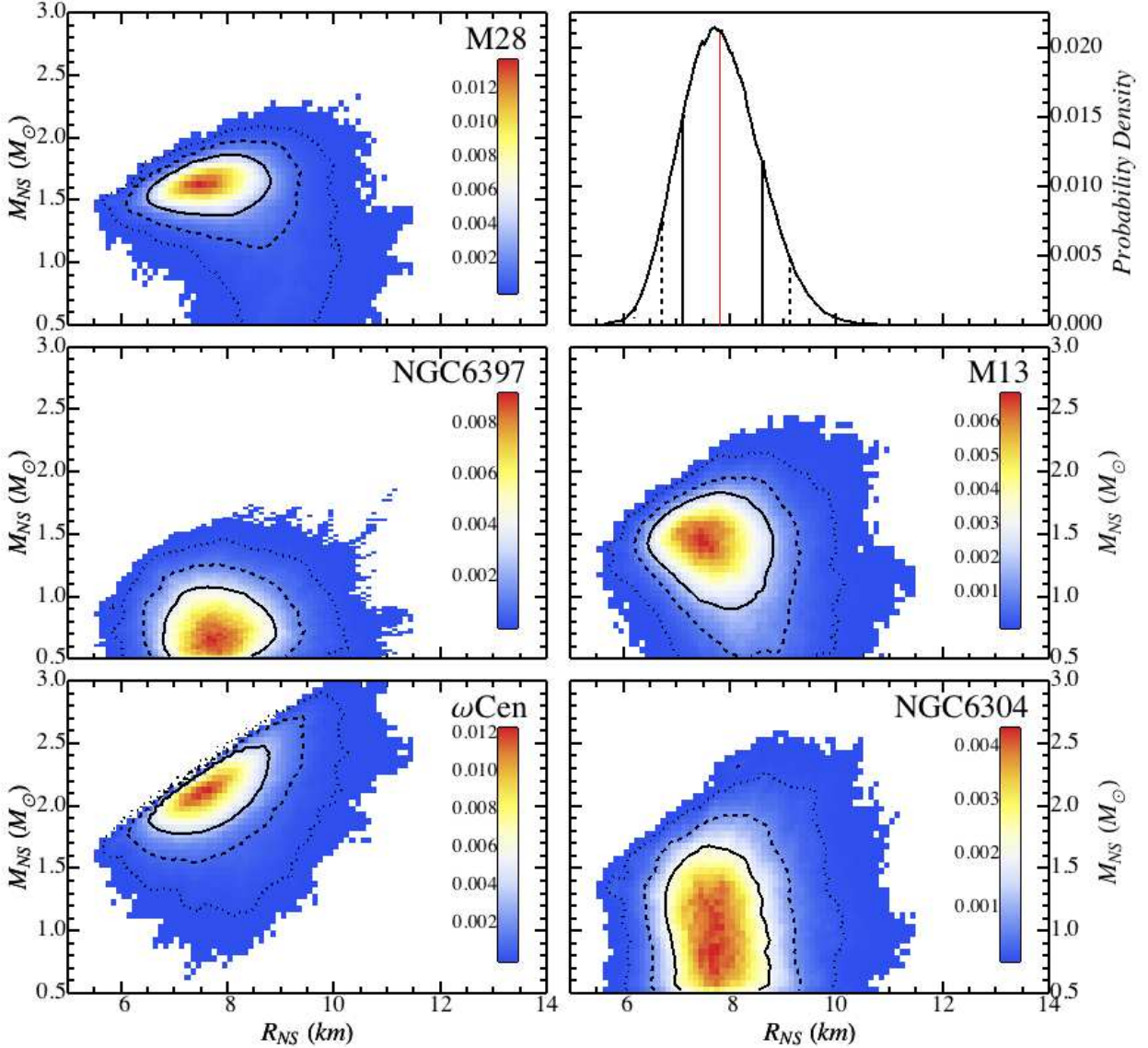,width=17cm,angle=0}~}
  \caption[]{\label{fig:FreeNhFreeDNoPL} Figure similar to
    Figure~\ref{fig:fixNhfixDNoPL}, but for the MCMC Run \#6.  The
    characteristics of this run include \nh\ values free to vary in
    the fit, and the presence of the Gaussian Bayesian priors for the
    distances.  No PL component was included in run.  A value
    $\rns=7.8\ud{1.3}{1.1}\km$ was found.}
\end{figure*}

\begin{figure*}[ht]
  \centerline{~\psfig{file=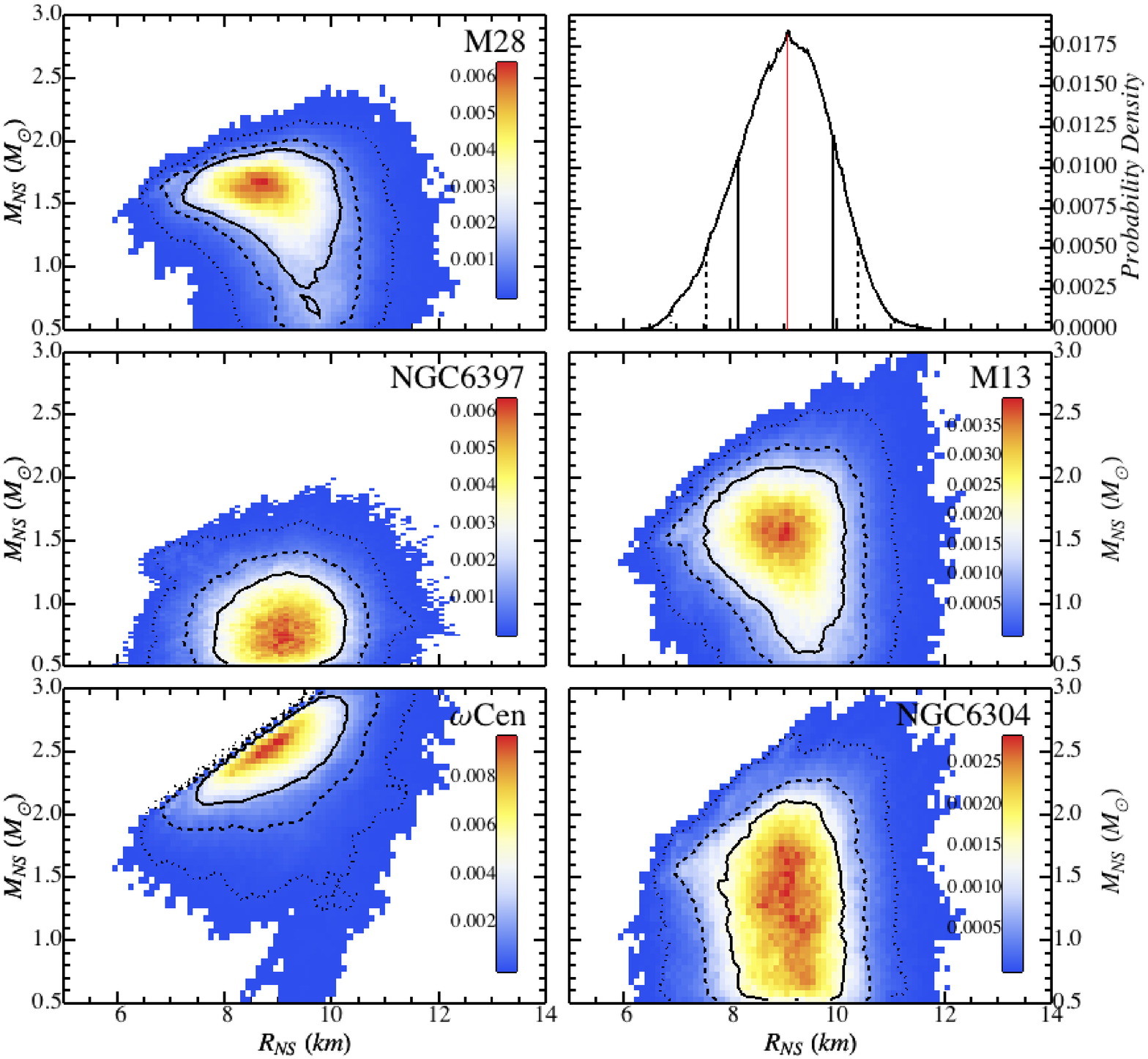,width=17cm,angle=0}~}
  \caption[]{\label{fig:FreeNhFreeDwithPL} Figure similar to
    Figure~\ref{fig:fixNhfixDNoPL}, corresponding to Run \#7.  Here,
    all the possible assumptions have been relaxed to obtain a
    \rns\ measurement the least affected by systematic uncertainties.
    The \nh\ parameters are left free; and Gaussian Bayesian priors
    and PL components are included.  This results in an
    \rns\ measurement: $\rns=9.1\ud{1.3}{1.5}\km$}
\end{figure*}

Following the first run, all assumptions (on distance, \nh\ and the
presence of a PL) are progressively relaxed in the MCMC Runs \#2
through \#7, with the last one producing the \rns\ measurement with
the fewest assumptions.  The \mr\ contours and \rns\ distribution of
each of these are displayed in
Figures~\ref{fig:fixNhfixDNoPL}--\ref{fig:FreeNhFreeDwithPL} and the
results are listed in Tables~\ref{tab:SimulFixNh} and
\ref{tab:SimulFreeNh}.  In the process of relaxing assumptions, one
confirms that the results remain consistent between each run and such
process does not significantly bias the results.  The effects of
relaxing each assumption are briefly described below.

\begin{deluxetable*}{ccccccc}
  \tablecaption{\label{tab:SimulFreeNh} Results from the simultaneous spectral fitting, with free \nh}
  \tablewidth{0pt}
  \tabletypesize{\scriptsize}    
  \tablecolumns{7}
  \tablehead{
    \colhead{Target} & \colhead{$\alpha_{\rm pileup}$} & 
    \colhead{\kteff} & \colhead{\mns} & \colhead{\rinfty} & \colhead{\nhtt} & \colhead{PL Norm $\tee{-7}$}\\
    \colhead{} & \colhead{} & \colhead{(eV)} & \colhead{(\Msun)} & 
    \colhead{(\km)} & \colhead{} & \colhead{$\PLunit$}}
  \startdata
   \multicolumn{7}{c}{Run~\#5: {\tt Free \nh, Fixed $d_{\rm GC}$, No PL included}, \rns=7.5\ud{1.1}{1.0}\km}\\
   \multicolumn{7}{c}{\Chisq{0.98}{638}{0.66}, 8\% accept. rate}\\  
   \hline
   M28      & 0.43\ud{0.13}{0.13} & 165\ud{24}{22} & 1.60\ud{0.21}{0.25}  & 12.4\ud{1.6}{1.5} & 0.248\ud{0.023}{0.022}  &  -- \\ 
   NGC~6397 &       --            &  70\ud{8}{5}   & 0.76\ud{0.33}{0.22p} &  9.0\ud{1.4}{1.1} & 0.105\ud{0.016}{0.015}  &  -- \\ 
   M13      &       --            & 106\ud{20}{16} & 1.43\ud{0.36}{0.44p} & 11.4\ud{2.5}{1.7} & 0.010\ud{0.024}{0.009p} &  -- \\
   \OmCen   &       --            & 137\ud{32}{31} & 2.00\ud{0.36}{0.41}  & 16.7\ud{4.7}{4.3} & 0.152\ud{0.048}{0.049}  &  -- \\ 
   NGC~6304 &       --            & 120\ud{28}{14} & 1.13\ud{0.59}{0.54p} & 10.1\ud{3.2}{1.9} & 0.315\ud{0.079}{0.060}  &  -- \\ 
   \hline
   \hline	
   \multicolumn{7}{c}{Run~\#6: {\tt Free \nh, Gaussian Bayesian priors for $d_{\rm GC}$, No PL included}, \rns=7.8\ud{1.3}{1.1}\km}\\
   \multicolumn{7}{c}{\Chisq{0.99}{633}{0.59}, 8\% accept. rate}\\  
   \hline
   M28      & 0.42\ud{0.13}{0.13} & 158\ud{26}{29} & 1.59\ud{0.25}{0.45p} & 12.4\ud{1.8}{1.7} & 0.248\ud{0.023}{0.022}  &  -- \\ 
   NGC~6397 &       --            &  71\ud{9}{5}   & 0.78\ud{0.39}{0.24p} &  9.4\ud{1.6}{1.3} & 0.104\ud{0.016}{0.015}  &  -- \\
   M13      &       --            & 100\ud{23}{18} & 1.38\ud{0.42}{0.64p} & 11.3\ud{2.6}{2.0} & 0.010\ud{0.023}{0.009p} &  -- \\ 
   \OmCen   &       --            & 133\ud{35}{30} & 2.07\ud{0.41p}{0.43} & 17.1\ud{5.2}{4.4} & 0.156\ud{0.050}{0.048}  &  -- \\ 
   NGC~6304 &       --            & 116\ud{30}{14} & 1.09\ud{0.68}{0.52p} & 10.3\ud{3.4}{1.9} & 0.321\ud{0.078}{0.061}  &  -- \\  
   \hline
   \hline	
   \multicolumn{7}{c}{Run~\#7: {\tt Free \nh, Gaussian Bayesian priors for $d_{\rm GC}$, PL included}, \rns=9.1\ud{1.3}{1.5}\km}\\  
   \multicolumn{7}{c}{\Chisq{0.98}{628}{0.64}, 7\% accept. rate}\\ 
   \hline
   M28      & 0.34\ud{0.14}{0.14} & 137\ud{29}{22} & 1.50\ud{0.37}{0.80p} & 12.6\ud{2.0}{2.0} & 0.248\ud{0.024}{0.023}  & 5.0\ud{3.7}{3.4p} \\ 
   NGC~6397 &       --            &  67\ud{8}{5}   & 0.86\ud{0.47}{0.31p} & 10.8\ud{1.7}{1.7} & 0.116\ud{0.017}{0.017}  & 2.7\ud{1.3}{1.3}  \\ 
   M13      &       --            &  92\ud{24}{15} & 1.47\ud{0.62}{0.78p} & 12.6\ud{3.7}{2.3} & 0.014\ud{0.028}{0.012p} & 2.4\ud{4.1}{2.1p} \\ 
   \OmCen   &       --            & 130\ud{33}{31} & 2.42\ud{0.42p}{0.54} & 20.3\ud{5.6}{5.7} & 0.172\ud{0.047}{0.051}  & 2.9\ud{3.9}{2.6p} \\ 
   NGC~6304 &       --            & 112\ud{31}{15} & 1.32\ud{0.80}{0.71p} & 12.1\ud{4.3}{2.5} & 0.346\ud{0.086}{0.065}  & 1.8\ud{2.6}{1.5p} \\ 
  \enddata

  \tablecomments{ This table is similar to Table~\ref{tab:SimulFixNh},
    abbreviations and symbols are the same.  The only difference is
    that the absorption \nh\ remained free for these runs. Quoted
    uncertainties are 90\% confidence as well. ``p'' indicates that
    the posterior distribution did not converge to zero probability
    within the hard limit of the model.}
\end{deluxetable*}

\begin{deluxetable*}{ccccccc}
  \tablecaption{\label{tab:nsagrav} Results from Simultaneous Spectral Fitting with {\tt nsagrav}}
  \tablewidth{0pt}
  \tabletypesize{\scriptsize}    
  \tablecolumns{7}
  \tablehead{
    \colhead{Target} & \colhead{$\alpha_{\rm pileup}$} & 
    \colhead{\kteff} & \colhead{\mns} & \colhead{\rinfty} & \colhead{\nhtt} & \colhead{PL Norm $\tee{-7}$}\\
    \colhead{} & \colhead{} & \colhead{(eV)} & \colhead{(\Msun)} & 
    \colhead{(\km)} & \colhead{} & \colhead{$\PLunit$}}
  \startdata
   \multicolumn{7}{c}{Run~\#8: {\tt Fixed \nh, Fixed $d_{\rm GC}$, No PL included}, \rns=7.8\ud{0.5}{0.3}\km}\\
   \multicolumn{7}{c}{\Chisq{1.04}{643}{0.25}, 10\% accept. rate} \\
   \hline
   M28      & 0.46\ud{0.11}{0.11} & 166\ud{4}{13}  & 2.03\ud{0.42p}{0.43}  & 16.3\ud{12.6}{3.9} & (0.252) & -- \\ 
   NGC~6397 &       --            &  66\ud{3}{2}   & 0.51\ud{0.23}{0.15p}  &  8.7\ud{0.6}{0.5}  & (0.096) & -- \\ 
   M13      &       --            & 101\ud{12}{11} & 1.42\ud{0.88p}{0.49p} & 11.5\ud{11.2}{1.7} & (0.008) & -- \\
   \OmCen   &       --            & 113\ud{2}{3}   & 1.86\ud{0.36p}{0.15}  & 14.3\ud{4.8}{1.2}  & (0.182) & -- \\ 
   NGC~6304 &       --            & 126\ud{17}{19} & 1.41\ud{0.97p}{0.74p} & 11.4\ud{14.1}{2.3} & (0.346) & -- \\  
  \enddata

  \tablecomments{ This table is similar to Table~\ref{tab:SimulFixNh}.
    But this run was performed with the {\tt nsagrav} model instead of
    the {\tt nsatmos} model for comparison purposes.  The differences
    between the two models are described in Section~\ref{sec:radius},
    for example, in the high-redshift regime (See
    Figure~\ref{fig:nsagrav}).  ``p'' indicates that the posterior
    distribution did not converge to zero probability within the hard
    limit of the model.  Quoted uncertainties are 90\% confidence.}
\end{deluxetable*}


\subsubsection{Using Gaussian Bayesian Priors for Distances}
When adding Gaussian Bayesian priors in place of the fixed distance
parameters, the \mr\ contours are, as expected, broader in the
\rinfty\ direction.  Because the normalization of a thermal spectrum
such a {\tt nsatmos} is $\propto \left( \rinfty / d \right) ^{2}$,
relaxing the assumptions on $d_{\rm GC}$ increases the possible values
of $\rinfty$.  This effect is mostly noticeable for the two targets
observed with the highest S/N, i.e., M28 and NGC~6397.  In Run \#2,
the posterior distribution of \rns\ corresponds to
$\rns=7.6\ud{0.9}{0.9}\km$, broader than that of the previous run.

\subsubsection{Adding a Power-Law Spectral Component}
When adding PL components to account for possible excesses of photons
at high energy, one finds NS parameters consistent with those of the
previous runs.  Most PL normalizations are consistent with zero in
Runs~\#3, \#4 and \#7.  For M28, the PL normalizations in these runs
is consistent with zero, within $2.4\sigma$.  For NGC~6397 in Runs~\#3
and \#4, the consistency with zero is only marginal, within
$2.8\sigma$.  Finally, in Run \#7, the PL normalization of NGC~6397,
${\rm Norm}_{\rm PL, NGC~6397}=2.7\ud{1.3}{1.3}\tee{-7}\PLunit$, is
not consistent with zero.  This may indicate the possible contribution
of a PL component at large energies and it is further discussed in
Section~\ref{sec:discussion}.

\subsubsection{Relaxing the \nh\ assumption}
As the constraints on the \nh\ parameters are relaxed, the {\tt
  nsatmos} best-fit parameters remain consistent with those of the
previous runs.  The posterior distributions of the five
\nh\ parameters are consistent with the \xray\ deduced \nh\ values
found from the spectral fits of the sources individually, i.e., the
values used in Run \#1--4.

\subsubsection{All Assumptions Relaxed}
For this final run (\#7), $\rns=9.1\ud{1.3}{1.5}\km$ is consistent
with the radii obtained in the previous MCMC runs (\#1--6).  The
posterior distributions of \mr\ are shown in
Figure~\ref{fig:FreeNhFreeDwithPL} and detailed in
Table~\ref{tab:SimulFreeNh}.  Once again, all resulting values are
consistent with those of the previous runs.  Progressively
relaxing assumptions ensures a good understanding of the spectral fit,
with no unexpected behavior.

\subsubsection{Comparison with {\tt nsagrav}}
We also performed the fit using the model {\tt nsagrav} instead of
{\tt nsatmos} for comparison purposes.  It has previously been shown
that {\tt nsatmos} and {\tt nsagrav} produce similar spectral
parameters when fit to experimental data \citep[e.g.][]{webb07}.  In
\emph{XSPEC}, for the {\tt nsagrav} model, the \rns\ range is
$\left[6\km,20\km\right]$, and the \mns\ range is
$\left[0.3\msun,2.5\msun\right]$ compared to
$\left[5\km,30\km\right]$ and $\left[0.5\msun,3.0\msun\right]$ with
     {\tt nsatmos}.  This run (\#8) was done with the same
     characteristics as Run \#1.

\begin{figure*}[ht]
  \centerline{~\psfig{file=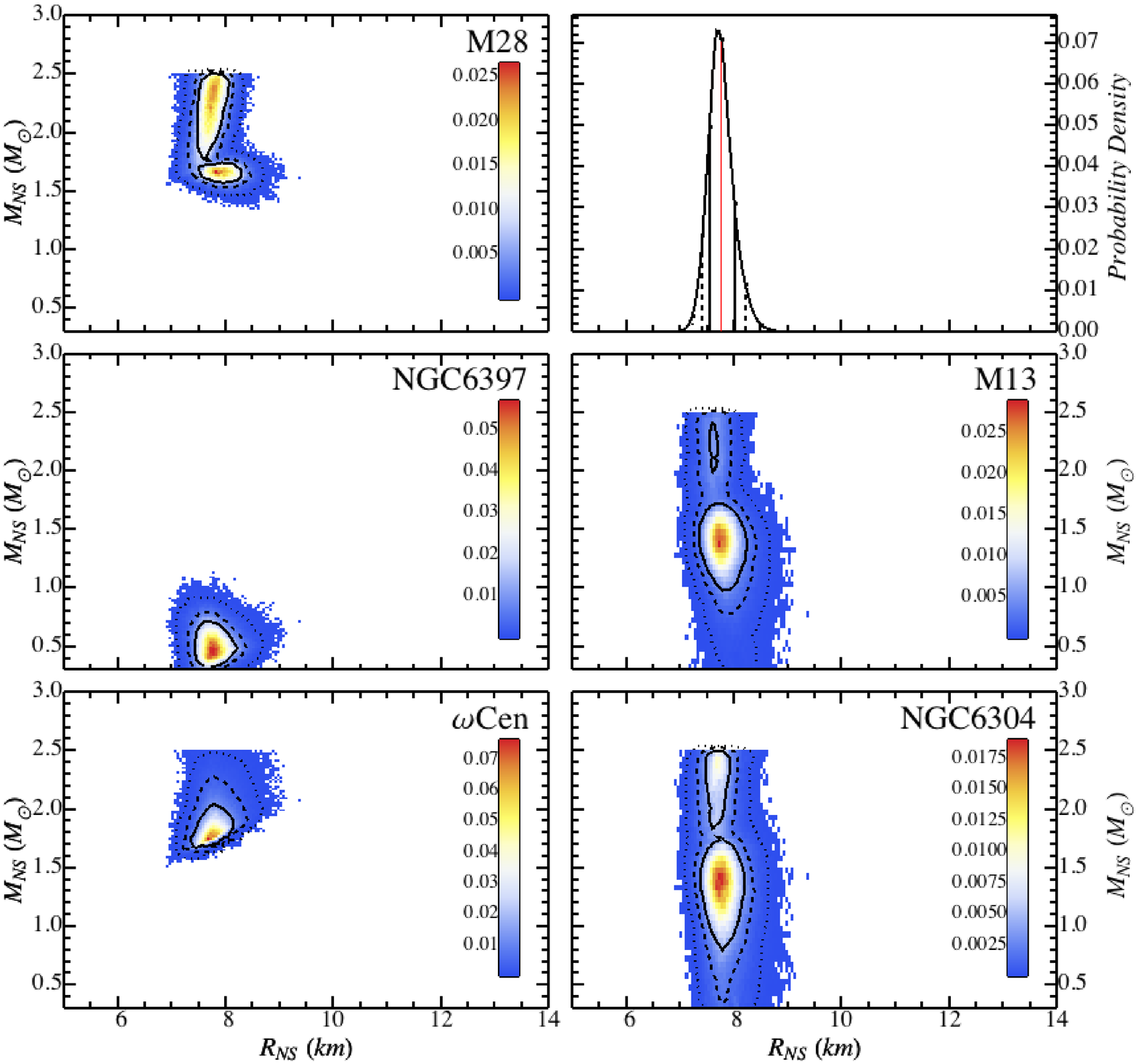,width=17cm,angle=0}~}
  \caption[]{\label{fig:nsagrav} Figure similar to
    Figure~\ref{fig:fixNhfixDNoPL}, i.e., Run \#1 with fixed distances
    instead of Gaussian Bayesian priors, fixed \nh\ values, and no PL
    component included.  The only difference resides in the
    H-atmosphere model, where {\tt nsagrav} model was used instead of
    {\tt nsatmos}. Note that the parameter limits in {\tt nsagrav} are
    [6--20]\km for \rns, and [0.4--2.5]\msun for \mns.  The
    \rns\ measurement, $\rns=7.8\ud{0.5}{0.3}\km$, is consistent with
    that of Run \#1, but the some of the \mr\ posterior distributions
    are significantly different from those of
    Figure~\ref{fig:fixNhfixDNoPL}. This is further discussed in
    Section~\ref{sec:radius}.}
\end{figure*}

When comparing the posterior distributions of the parameters and
\mr\ contours obtained with {\tt nsagrav} (Figure~\ref{fig:nsagrav}
for Run \#1 and Table~\ref{tab:nsagrav}) to those obtained with {\tt
  nsatmos} (Figure~\ref{fig:fixNhfixDNoPL} and
Table~\ref{tab:SimulFixNh}), some consistencies can be noticed.
However, not all distributions are consistent between the two models.
Specifically, for M28, M13, and NGC 6304, one can notice that an
additional distinct lobe at high \mns\ appears in the \mr\ parameter
space. This appears to be because the {\tt nsagrav} model as
implemented in {\tt XSPEC} gives different values in this parameter
space than returned by {\tt nsatmos}; the authors of this model state
that this is because the model is inapplicable in this parameter
region\footnote{The \mr\ space where {\tt nsagrav} is applicable can
  be seen here
  \url{http://heasarc.gsfc.nasa.gov/xanadu/xspec/models/m-r.pdf}}
(Zavlin and Pavlov, priv. comm.).  For example, some sets of
\mr\ allowed by {\tt nsagrav} and giving an acceptable \chisq-value
lead to imaginary values of \rinfty.  It is important for an observer
to keep this fact in mind, otherwise, results produced by the {\tt
  XSPEC} implementation of {\tt nsagrav} could be misinterpreted.  In
light of the pitfall mentioned here, the {\tt nsagrav} model should be
used with care.

\section{Discussion}
\label{sec:discussion}
This paper presented the simultaneous analysis of the spectra from
five qLMXBs in GCs with a common \rns\ parameter for all targets. The
posterior distributions for \rns, \mns, \rinfty, \kteff, and \nh\ were
obtained from MCMC simulations, which included Gaussian Bayesian
priors for the distances to the GCs hosting the targets.  In this
discussion section, the original work performed here and the data used
are summarized.  This is followed by a subsection discussing various
possible biases resulting from the MCMC analysis.  The discussion
finishes with the implication that the resulting \rns\ measurement may
have for the determination of the dEoS.

\subsection{List of New Analysis Methods, Data and Results}
The following two paragraphs aim at summarizing the novel approach to
the analysis of the NS thermal spectra, and unused data presented in
this paper.  The MCMC framework for spectral analysis is a rather
recent approach, made convenient by the development of \emph{PyXSPEC},
the Python version of \emph{XSPEC}.  The ``Stretch-Move'' MCMC
algorithm used here differs from the usual Metropolis-Hasting (M-H)
algorithm of standard MCMC simulations.  It has been developed
recently and while it presents significant advantages over M-H, it is
still scarcely used in astrophysics.  This work includes
\xray\ data not previously presented in the literature, namely the
\chandra\ exposure of NGC~6304.

This paper also contains a more complete presentation of the resulting
best-fit NSs physical parameters (\kteff, \mns, and \rns).
Specifically, the products of the MCMC simulations in 2-dimensional
matrices of posterior distributions (see
Figures~\ref{fig:M28contours}--\ref{fig:NGC6304contours}) are
displayed.  Such figures better represent the true distributions of
the NS physical parameters compared to simple lists of best-fit values
with their uncertainties, mostly because the distribution are not
necessarily Gaussian (particularly with models like {\tt nsatmos}).
We encourage researchers to present their results in such a way.

Finally, the approach of this paper is not dissimilar to the work of
\cite{steiner10,steiner12} in the sense that different targets are
combined to produce constraints on the dEoS.  However, our analysis
imposes \rns\ to be quasi-constant for all the targets, as justified
by recent observations favoring ``normal matter'' dEoSs.  Other
differences include two qLMXBs (M28 and NGC~6304) added to the present
analysis, and the qLMXB in 47~Tuc not used here because of the
uncertainties related to pile-up.  Finally, the work of
\cite{steiner10,steiner12} uses type-I \xray\ burst sources, which are
not considered in the present analysis.

\subsection{Possible biases resulting from the analysis}
\subsubsection{Non-zero power-law for NGC~6397}
The runs that included a PL component (\#3, \#4, and \#7, with photon
index $\Gamma=1$) resulted in PL normalizations consistent with zero
for all targets, except for NGC~6397.  Specifically, ${\rm Norm}_{\rm
  PL,NGC~6397} = 2.7\ud{1.3}{1.3}\tee{-7}\PLunit$, which is not
consistent with zero, at the $3.4\sigma$ level, in Run \#7.  This
indicates a possible non-negligible contribution of a PL above $\sim
2\keV$.  Using \emph{XSPEC} (without the MCMC approach), the spectra of
NGC~6397 are fitted without the other qLMXBs, and the PL normalization
obtained is ${\rm Norm}_{\rm PL} = 2.1\ud{1.3}{1.3}\tee{-7}\PLunit$
which corresponds to a contribution of $2.6\ud{1.7}{1.7}\%$ of the
total unabsorbed flux of NGC~6397 in the 0.5--10\keV\ energy band.
Such a contribution is consistent with that measured in the previous
results, $\leq 3.3\%$ \citep{guillot11a}, for the same photon index.

Nonetheless, adding a PL contribution for each target does not
significantly bias the posterior distribution of \rns.  Specifically,
adding PL components between Runs \#1 and \#3 changed the radius
measurement from $\rns=7.1\ud{0.5}{0.6}\km$ (Run \#1) to
$\rns=7.3\ud{0.5}{0.6}\km$ (Run \#3); between Runs \#2 and \#4,
\rns\ changed from $\rns=7.6\ud{0.9}{0.9}\km$ (Run \#2) to
$\rns=8.0\ud{1.0}{1.0}\km$ (Run \#4).  Between runs \#6 and \#7 (free
\nh), the \rns\ distributions changed more significantly, but they are
nonetheless consistent: $\rns=7.8\ud{1.3}{1.1}\km$ (Run \#6) to
$\rns=9.1\ud{1.3}{1.5}\km$ (Run \#7).  Therefore, adding PL components
does not significantly change the \rns\ posterior distribution, but
nonetheless includes systematic uncertainties related to the possible
presence of a PL component into the measured \rns.  With the limited PL
contributions observed ($<5\%$), the choice of photon index does not
affect the {\tt nsatmos} component.  This is tested on NGC~6397 alone,
the qLMXB with the strongest PL contribution, where the PL photon
index is changed from $\Gamma=1$ to $\Gamma=2$.  Such a change results
in consistent PL contributions, but more importantly, it results in a
non-significant increase of only $\sim1\%$ in the value of \rns, well
within the uncertainties of the measurement.  Other parameters such as
the mass, temperature and the PL normalization are consistent between
the two trials, with $\Gamma=1$ and with $\Gamma=2$.  Given that
NGC~6397 has the strongest PL contribution and that changing the
photon index from $\Gamma=1$ to $\Gamma=2$ does not modify \rns, it is
expected to be the same for the simultaneous spectral analysis.

\subsubsection{Effects of Individual Targets on the Simultaneous Fit}
\label{sec:add47tuc}

An additional set of MCMC runs was performed to investigate whether
some targets have a dominant biasing effect on the simultaneous
spectral fitting presented above.  This analysis is performed by
excluding each target individually and observing the resulting
marginalized posterior distributions of the parameters.  Results are
listed in Table~\ref{tab:WithoutIndividual}.  These runs were
performed like Run \#7, i.e., with the \nh\ value free, with Gaussian
Bayesian priors on the distances, and with the additional PL
component.  The values of \rns\ obtained when removing each target are
all consistent within $2\sigma$ of each other, and more importantly,
consistent with the \rns\ distribution obtained when all targets are
included (Run \#7, in Table~\ref{tab:SimulFreeNh}).  Similarly, the
\kteff, \mns, \rinfty\ and the $\alpha_{\rm pileup}$ values of each
targets are consistent with each other in all five cases presented in
Table~\ref{tab:WithoutIndividual}, and also consistent with the values
from Run \#7.  Overall, no individual target has any dominant effect
on the simultaneous fit.  This ensures that no qLMXB induces a
significant bias on \rns.  This is also compatible with recent results
\citep{steiner12}, obtained by combining \mr\ distributions of qLMXBs
and type-I \xray\ bursts.  In that work, it was demonstrated that
removing extreme cases (for example the qLMXB X7 in 47Tuc, or the
qLMXB in M13) had no or little effect on the resulting empirical dEoS.

For completeness, and to confirm this observation, a spectral analysis
combining X7 to the other five qLMXBs of this work is performed in
\emph{XSPEC} in order to determine $\rns$.  Note that the MCMC
approach is not used and is not necessary here since this additional
analysis is simply a consistency check to determine if outliers can
have an impact of the best-fit \rns measurement\footnote{As mentioned
  before, X7 was not used in the main analysis of the present work
  because of the unquantified systematic errors associated with the
  correction of the large amount of pile-up affecting the spectra, and
  its uncertain effect on our final statistical error bars.}.
Furthermore, Gaussian Bayesian priors are not used for this
consistency check. The spectral data available for X7 and used here
are described in a previous work \citep{heinke06a}.  For the distance
to the qLMXB, we used the weighted average of all recent distance
estimates listed in another reference \citep{woodley12}, i.e., $d_{\rm
  47Tuc}=4.52\kpc$.  The resulting best-fit NS radius found was
$\rns=7.0\ud{1.0}{2.0p}\km$ with the acceptable statistics
\Chisq{0.98}{1003}{0.70}.  This result is consistent with the
\rns\ measurement of Run \#1 (performed without adding the spectra of
X7) and seem to confirm that outliers, such as the qLMXB X7 in 47Tuc,
do not affect the radius measurement, as demonstrated in a previous
work \citep{steiner12}.

\subsubsection{Composition of the Neutron Star Atmosphere}
The \mr\ measurements of NSs in qLMXBs rely on the atmosphere
modeling, which itself relies on a major assumption of this work,
namely the composition of the NS atmosphere.  It is generally assumed
that NS atmosphere are composed exclusively of pure hydrogen
\citep[e.g.,][]{rutledge02b,heinke06a}, since heavier accreted
elements will settle through the atmosphere on short time scales
\citep{bildsten92}. NS He atmospheres could be observed in
the case of He accretion from WD donors in ultra-compact
\xray\ binaries.  An He-atmosphere model was used to fit the spectra
of M28 and led to a radius $1.5\times$ larger than that obtained with a
H-atmosphere, favoring stiff dEoSs \citep{servillat12}.  Similarly,
for the qLMXB in M13, assuming a He composition of the NS
atmosphere increases the radius \rns\ by a factor $\sim1.2$
\citep{catuneanu13}.

The actual composition of the NS atmosphere can be infered from the
identification of the donor companion star, which proves a difficult
task in the crowded environment of GCs.  Only two GC qLMXBs have
identified counterparts, X5 in 47Tuc \citep{edmonds02}, and in
\OmCen\ \citep{haggard04}, both of which discovered from their strong
${\rm H\alpha}$ emission, indicating hydrogen donor stars.

\subsubsection{Causality Limit}
With the assumptions made and the model chosen in the present work,
parts of the \mr\ contours resulting from the analysis cover a section
of the parameter space that goes past the causality limit as set in
ealier works \citep{lattimer90,lattimer01}.  Stricter constraints on
the \mr\ contours could be obtained by imposing that the sets of MCMC
accepted points (\rns,\mns) all obey $\rns\geq 3G\mns/c^2$, i.e., do
not cross the causality line.  However, we choose not to change
constraints; this produces a larger error region than may be necessary
(if one were to explicitly adopt causality as an assumption), but the
goal of this analysis is to produce the most conservative,
assumption-free uncertainty region for \rns.  Only if much higher S/N
data were obtained, and the M-R parameter space required (or strongly
preferred) a value in the region excluded by the causality requirement
cited above \citep{lattimer90} would it become necessary to revisit
this assumption.

\begin{deluxetable}{cccc}
  \tablecaption{\label{tab:WithoutIndividual} Effect of Individual Targets on
    the Simultaneous Spectral Fit}  
  \tablewidth{0pt}
  \tabletypesize{\scriptsize}    
  \tablecolumns{4}
  \tablehead{
    \colhead{Target}    & \colhead{\rns} & \colhead{$\chisqnu$/d.o.f. (prob.)} & \colhead{Accept.} \\
    \colhead{excluded} & \colhead{(\km)} & \colhead{} & \colhead{rate} }
  \startdata
    NONE (Run~\#7)   &  9.1\ud{1.3}{1.5}\km & 0.98 / 628 (0.64) & 7\%\\
    \hline
    WITHOUT M28      &  8.4\ud{1.5}{1.3}\km & 0.98 / 381 (0.69) & 7\%\\
    WITHOUT NGC~6397 & 10.7\ud{1.7}{1.4}\km & 0.89 / 428 (0.95) & 9\%\\
    WITHOUT M13      &  8.6\ud{1.5}{1.3}\km & 0.94 / 588 (0.86) & 7\%\\
    WITHOUT \OmCen   &  8.7\ud{1.5}{1.4}\km & 0.95 / 601 (0.81) & 8\%\\
    WITHOUT NGC~6304 &  9.0\ud{1.5}{1.4}\km & 0.93 / 622 (0.88) & 8\%\\
  \enddata

  \tablecomments{ The spectral fits in this table were performed
    following Run \#7, with free \nh\ values, Gaussian Bayesian priors
    on the distances and with an additional PL components in the
    model.  Each target were successively removed to investigate the
    possible effect of individual sources on the global fit.  The
    \rns\ values obtained in each of these five tests are consistent
    with each other, and with that of Run \#7.  This confirms that
    none of the five qLMXBs significantly skews the $\rns$
    measurement.  Quoted uncertainties are 90\% confidence, and values
    in parentheses are held fixed.}
\end{deluxetable}

\subsubsection{Effect of Assumptions}
As discussed in this paper, assumptions can have a strong effect on
the interpretation of spectral fits of individual sources and
consequently on the simultaneous spectral analysis as well.  The
selection of the distances to the GCs (fixed or with Gaussian Bayesian
priors) can skew the \rns\ measurement toward smaller or larger
values. For instance, in early runs of this analysis, the distance
$d_{\rm M28}=5.1\pm0.5\kpc$ \citep{rees91} was initially used, before
it was updated to a more recent value $d_{\rm M28}=5.5\pm0.3\kpc$
\citep{testa01,servillat12}.  This caused the \rns\ measurement (of Run \#7)
to change from $\rns=8.7\ud{1.3}{1.1}\km$ to
$\rns=9.1\ud{1.3}{1.5}\km$ as the distance of M28 was increased.  The
values are consistent with each other, but this larger $d_{\rm M28}$
shifted the \rns\ measurement. This is expected since the
normalization of a thermal spectrum is $\propto \left( \rinfty / d
\right) ^{2}$.

Another strongly influential parameter is the galactic absorption.  A
change in \nh\ will strongly affect the \mns\ and \rns\ best-fit.
Specifically, a decrease (increase) in \nh\ results in an decrease
(increase) in \rinfty, respectively. This was shown for NGC~6397 and
\OmCen, in Section~\ref{sec:rinfty}.  For NGC~6397, the HI-survey
\nh\ value ($\nhtt=0.14$) leads to a $\rinfty=11.8\ud{0.8}{0.7}\km$,
while the \xray\ deduced value ($\nhtt=0.096\ud{0.017}{0.015}$)
produces $\rinfty=8.4\ud{1.3}{1.1}\km$, a decrease of $\sim30\%$.  For
\OmCen, the best-fit \rinfty\ almost doubles as the constraint on
\nh\ is relaxed, causing the value to increase from $\nhtt=0.09$
(HI-survey) to $\nhtt=0.182\ud{0.041}{0.047}$.  Consequently, using
assumptions for the \nh\ values, like those derived from HI-surveys,
can lead to strongly skewed \rns\ measurements.

Consequently, it is preferable to avoid using assumed values, when
possible, and therefore the \rns\ measurement from Run \#7 is
presented as the final result of this work.

\subsection{The \rns\ measurement} 

\begin{figure*}[ht]
  \centerline{~\psfig{file=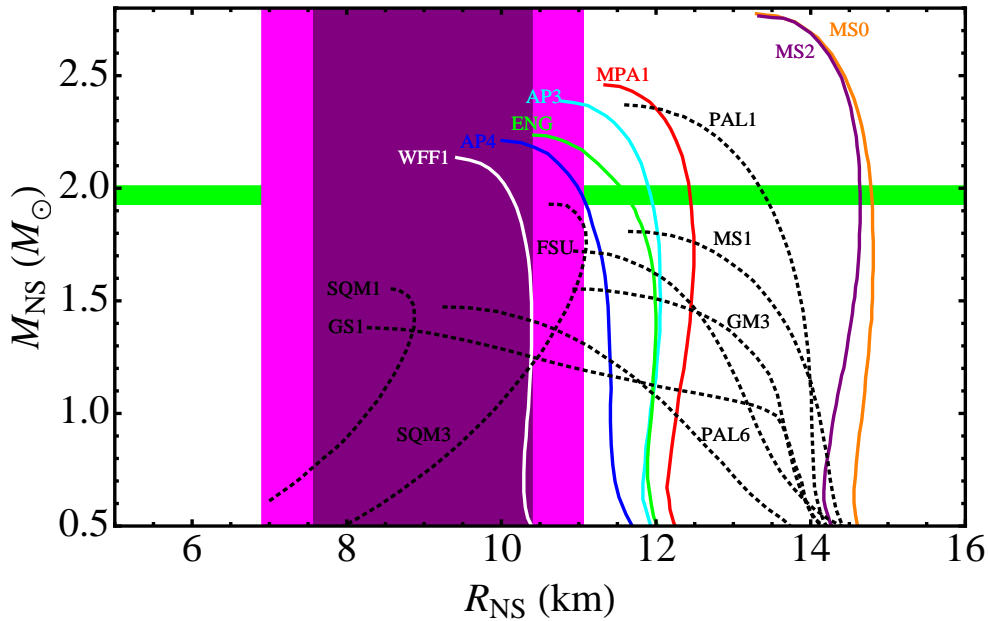,width=13cm,angle=0}~} 
  \caption[]{\label{fig:eos} Figure showing the constraint on the dEoS
    imposed by the radius measurement obtained in this work:
    $\rns=9.1\ud{1.3}{1.5}\km$ (90\%-confidence). The dark and light
    shaded areas show the 90\%-confidence and 99\%-confidence
    constraints of the \rns\ measurement, respectively.  The mass
    measurement of PSR~J1614-2230 is shown as the horizontal band
    \citep{demorest10}.  ``Normal matter'' EoSs are the colored solid
    lines.  Other types of EoSs, such as the hybrid or quark-matter
    EoSs are included for comparison, with dashed lines.  As mentioned
    in Section~\ref{sec:discussion}, the present analysis only places
    constraints on the ``normal matter'' EoSs since they are the only
    family of EoSs included in our assumptions.  Among them, only the
    very soft dEoSs (such as WFF1, \citealt{wiringa88}) are consistent
    with the radius obtained here.  The EoS are obtained from
    \cite{lattimer01,lattimer07}.}
\end{figure*}

The goal of this paper being to measure \rns, the following discussion
is focused on the \rns\ posterior distributions, on the comparison
with other \rns\ measurement, and on the implication that such a
radius measurement will have for the determination of the dEoS.

The striking observation one can make pertains to the low range of
values of the resulting \rns\ distributions obtained from the
different runs.  The \rns\ distributions remained below $\rns<10.4\km$
(90\%-confidence), or $\rns<11.1\km$ (99\%-confidence).  This resulted
from Run \#7 (see Figure~\ref{fig:FreeNhFreeDwithPL}), where a
particular effort was made to consider all possible sources of
systematic uncertainties.  The result from Run \#7 is the most general
\rns\ distribution, i.e., with the fewest assumptions, that can be
produced.  Also, the progressive relaxation of the assumptions
throughout the analysis demonstrated that no unexpected behavior was
present in the final \mr\ distributions of Run \#7 and that the
resulting low-value of \rns\ was not affected by systematics.

Previous works reported low values of NS radii, but these
measurements have high uncertainties due to low S/N, leading to poorly
constrained \rns\ and \mns\ (e.g., in NGC~2808, 
\citealt{webb07,servillat08}).  Another qLMXB in NGC~6553 was
identified with a small radius, $\rns=6.3\ud{2.3}{0.8}\km$
(90\%-confidence) for $\mns=1.4\msun$ \citep{guillot11b}.  However,
low-S/N \chandra\ observations demonstrated that the \xmm\ spectra of
the source was affected by hard \xray\ contamination from a marginally
resolved nearby source.  Higher-S/N observations with \chandra\ are
necessary to confirm the qLMXB classification and produce the
uncontaminated spectrum necessary for its use in the present analysis.

In addition to qLMXB \rns\ measurements, low radii were found from the
analysis of photospheric radius expansion type-I \xray\ bursts.  A
review of the method used to determine \rns\ from these sources can be
found in the literature \citep{ozel06,suleimanov11a}.  The LMXBs
EXO~1745-248, 4U~1608-52, and 4U~1820-30 were found to have respective
radii in the $2\sigma$ ranges $\rns=[7.5-11.0]\km$ \citep{ozel09a},
$\rns=[7.5-11.5]\km$ \citep{guver10a} and $\rns=[8.5-9.5]\km$
\citep{guver10b}, respectively.  While these results are on a par with
what is found in this paper, controversy emerged with the realization
that the analysis presented in the cited works was not internally
consistent because the most probable observables (from Monte-Carlo
sampling) led to imaginary masses and radii \citep{steiner10}.
Relaxing the assumption that the photospheric radius equals the
physical radius \rns\ at touchdown led to real-valued solutions of
\mns\ and \rns, and to larger upper limits for the
radius. Furthermore, it is argued in a later work that the short
bursts from EXO~1745-248, 4U~1608-52 and 4U~1820-30 are not
appropriate for such analysis because the post-burst cooling evolution
of these sources does not match the theory of passively cooling NSs
\citep{suleimanov11b}.  Therefore, the \mr\ constraints from type I
\xray\ bursts should be considered with these results in mind.

More recently, distance independent constraints in \mr\ space were
produced from the analysis of the sub-Eddington \xray\ bursts from the
type I \xray\ burster GS~1826-24 \citep{zamfir12}.  That analysis,
performed for a range of surface gravities
($\log_{10}\left(g\right)=14.0,14.3,14.6$) and a range of H/He
abundances ($0.01\unit{Z_{\odot}}$, $0.1\unit{Z_{\odot}}$ and
$Z_{\odot}$) led to radii $\rns\simlt11.5\km$.  While
distance-independent, the results are highly influenced by the
atmosphere composition and metallicity.  For pure He composition, the
upper limit of \rns\ becomes $\rns\simlt15.5\km$ \citep{zamfir12}.

Finally, the multiwavelength spectral energy distribution of the
isolated neutron star RX~J185635-3754 was analyzed to produce small
values of \rns\ and \mns\ with no plausible dEoS consistent with these
values: $\rns\sim6\km$ and $\mns\sim0.9\msun$ for $d=61\pc$
\citep{pons02}.  A recent distance estimation to the source
$d=123\ud{11}{15}\pc$ \citep{walter10} led to revised values:
$\rns=11.5\ppm1.2\km$ and $\mns=1.7\ppm1.3\msun$ \citep{steiner12}.
While this result is consistent with the \rns\ measurement obtained in
this paper and with the other works reporting low-\rns\ values, it has
to be taken with care since the high-magnetic field of the source is
not accounted for in the spectral model used by the original analysis.

Recently, it was shown that the dEoS can be empirically determined
from \mr\ measurements of NS, using the thermal spectra of qLMXBs and
the photospheric radius expansion of \xray\ bursts \citep{steiner10}.
This method uses MCMC simulation and Bayesian priors to determine the
most probable dEoS parameters, and equivalently, the corresponding
most probable \MofR\ for NS.  In a recent paper, this method was used
with four \xray\ bursting sources and four GC qLMXBs.  Considering all
scenarios, the $2\sigma$ lower and upper limits for \rns\ are
9.17\km\ and 13.92\km\ \citep{steiner12}.  The \rns\ distribution of
the present paper $\rns=9.1\ud{1.3}{1.5}\km$ (90\%-confidence, from
Run \#7) is consistent with several of the model variations of
\cite{steiner12}, namely variation C (dEoS parameterized with uniform
prior in the pressure at four energy density values), variation CII
(same as previous, but with low value of the color correction,
$1<f_{C}<1.35$), variation AII/AIII (dEoS parameterized as two
piecewise continuous power-laws, with $1<f_{C}<1.35$), see
\cite{steiner12} for details about the variations of the model.
Variation E (dEoS for quark stars) is incompatible with our original
assumption that \rns\ is quasi-constant for a large range of
\mns\ above 0.5\msun.  

Theoretical EoSs have been proposed for more than two decades.  A
non-exhaustive list can be found in the literature
\citep{lattimer01,lattimer07}.  When comparing the resulting
\rns\ distribution to proposed theoretical ``normal matter'' dEoSs,
one can note that most of those are not consistent with the
low-\rns\ result presented in this work.  Indeed, most of the dEoSs
describing ``normal matter'' correspond to radii larger than
11.5\km\ (see Figure~\ref{fig:eos}).  A spread in \rns\ ia observed in
these dEoSs at large masses, in the part of the \mr\ diagram where the
compact object approaches collapse.  However, this breadth of the
\rns\ variation for a given dEoS is well within the uncertainties
obtained in this work.  Overall, the radius measurement
$\rns=9.1\ud{1.3}{1.5}\km$ constrains the dEoS to those consistent
with low-\rns, such as WFF1 \citep{wiringa88}.  Note that this
analysis cannot address the veracity of more exotic types of EoSs
(hybrid and SQM) or any dEoS which does not predict a quasi-constant
\rns\ within the observable mass range.

It is known that \rns\ is related to fundamental nuclear physics
parameters, such as the symmetry energy
\citep{horowitz01a,horowitz01b}.  We expect the present constraints on
\rns\ can be used to constrain this, and other properties of dense
nuclear matter.  We leave this for future work.

It has been pointed out \citep{lattimer10} that an argument regarding a
maximally compact neutron star \citep{koranda97} results in a
relationship between the maximal neutron star radius ($R_{\rm max}$)
and the maximal neutron star mass ($M_{\rm max}$) for a given equation
of state:
\begin{equation}
\frac{R_{\rm max}}{M_{\rm max}} = 2.824 \frac{G}{c^2}
\end{equation}
\noindent where $G$ is Newton's constant and $c$ is the speed of
light.  Adopting this, the 99\% confidence upper limit is $M_{\rm
  max}<2.66 \msun$, which does not violate any measured neutron star
masses at present.

The small-\rns\ value found in this paper, and other low-\rns\ results
cited above, are consistent with soft dEoSs such as WFF1.  However,
results of \mns\ and \rns\ measurements from other sources seem to
favor stiffer dEoSs.  The qLMXB 47Tuc~X7 has a reported radius
$\rns=14.5\ud{1.6}{1.4}\km$ for $\mns=1.4\msun$ \citep{heinke06a},
supporting stiff dEoSs, such as MS0/2 \citep{muller96}.  Nonetheless,
the range of radii allowed by the published \mr\ contours for X7 is
consistent with the radius measurement presented in the present work.
Moreover, the X7 \mr\ contours are compatible with the dEoS WFF1
\citep{wiringa88}.  Another work used the long photospheric radius
expansion \xray\ bursts from 4U~1724-307 to conclude that stiff dEoSs
are describing the dense matter inside NSs \citep{suleimanov11b}.
Specifically, it was found that $\rns>13.5\km$ for $\mns<2.3\msun$,
and for a range for NS atmospheric composition.  Lower $\rns$ values,
in the range 10.5--17\km, are allowed for $\mns>2.3\msun$, for pure H
or solar metallicity composition.  This radius measurement is only
marginally consistent with the present work for large masses,
$\mns>2.3\msun$, which implies a dEoS capable of reaching
$\mns\sim2.3\msun$ for $\rns\sim10-11\km$.  Finally, another radius
measurement, obtained by modeling the thermal pulses of the
millisecond pulsar PSR~J0437$-$4715 \citep{bogdanov12}, led to values,
$\rns>11\km$ ($3\sigma$), is inconsistent with the measurement
presented in our work.

\section{Summary}
\label{sec:conclusion}
In this paper, we measured \rns\ using the assumption that the radius
is quasi-constant for a wide range of \mns\ larger than
$\mns>0.5\msun$, i.e., constant within the measurement precision.
This is justified by recent observations favoring ``normal matter''
dEoSs which are described by this characteristic.  For this analysis,
the spectra from five GCs qLMXBs observed with the \chandralong\ and
\xmmlong\ were used in a simultaneous analysis, constraining \rns\ to
be the same for all targets.

For this, we used an MCMC approach to spectral fitting, which offers
several advantages over the Levenberg-Marquardt
$\chi^{2}$-minimization technique generally used for spectral fits.
For example, the MCMC framework allows imposing Bayesian priors to
parameters, namely the distance to the host GCs.  By doing do, the
distance uncertainties are included into the posterior
\rns\ distribution.  In addition, one can marginalize the posterior
distributions over any parameters and very easily obtain
\mr\ distributions, while the grid-search method in \emph{XSPEC} can
be problematic in the case of spectral fits with many free parameters
and complicated \chisq-space.  The algorithm chosen in this work is an
affine-invariant ensemble sampler, commonly called ``Stretch-Move''
algorithm, which is particularly appropriate (i.e., converging
efficiently) for elongated and curved distributions.

The principal result of the simulations performed in this analysis is
that NSs are characterized by small physical radii.  Specifically,
when the distances and Galactic absorption parameters are fixed,
$\rns=7.1\ud{0.5}{0.6}\km$ (from Run \#1).  A more general posterior
distribution for \rns\, i.e., less prone to systematic biases, is
obtained by applying Gaussian Bayesian priors for the five GC
distance, by freeing the \nh\ parameters, and by adding a PL component
to the model to account for a possible spectral component at high
photon energies.  Such a spectral component could be the largest
possible source of uncertainty, and could be skewing \rns\ downward,
but it is accounted for in the last and most relaxed MCMC run.  In
fact, such a spectral component was discovered herein for NGC~6397.

The progressive relaxation of assumptions led us to a good
understanding of the spectral fit in Run \#7, minimizing systematic
uncertainties.  Therefore, with the H-atmosphere model {\tt nsatmos},
the measured NS radius is $\rns=9.1\ud{1.3}{1.5}\km$ (from Run \#7).
These results are compatible with other low-\rns\ measurements from GC
qLMXBs or type-I \xray\ bursts, but not consistent with some
published \rns\ measurement leading to values $\rns>11\km$.  We
recommend these \rns\ constraints, from Run \#7, be those relied upon
for constraints on the dEoS and other nuclear physics model
parameters, as this run has the fewest associated assumptions behind
it.

Among the dEoS listed in previous works \citep{lattimer01,lattimer07},
the \rns\ measurement presented here is only compatible with ``normal
matter'' dEoSs consistent with $\rns\sim10\km$, e.g. WFF1
\citep{wiringa88}.  Most dEoSs are compatible with larger radii, at
$\rns\approxgt12\km$ and above.  Given the results presented in this
work, the theory of dense nuclear matter may need to be revisited.

\acknowledgements The authors would like to thank the referee for
useful remarks that improved the clarity of this article.  SG is a
Vanier Canada Graduate Scholar and acknowledges the support of NSERC
via the Vanier CGS program.  RER is supported by an NSERC Discovery
grant.  MS acknowledges supports from NASA/Chandra grant GO0-11063X
and the Centre National d'\'{E}tudes Spatiales (CNES).  The authors
would like to thank Keith A. Arnaud and Craig Gordon for their
precious help with the use of \emph{XSPEC} and \emph{PyXSPEC}.  The
authors are also very grateful toward Ren{\'e} Breton for sharing his
python implementation of the ``Stretch-Move'' algorithm.  Finally, the
authors also acknowledge the use of archived \xmm\ and \Chandra\ data
from the High Energy Astrophysics Archive Research Center Online
Service, provided by the NASA GSFC.

\bibliographystyle{apj_8}
\bibliography{biblio}



\end{document}